\documentclass[acmsmall, screen]{acmart}

\acmJournal{POMACS}
\acmYear{2026}

\usepackage{booktabs}
\usepackage{tikz}
\usetikzlibrary{arrows.meta, positioning, fit}
\graphicspath{{fig/}}
\definecolor{c0}{RGB}{31,119,180}\definecolor{c1}{RGB}{214,39,40}\definecolor{c2}{RGB}{44,160,44}
\definecolor{c3}{RGB}{148,103,189}\definecolor{c4}{RGB}{255,127,14}
\usepackage{pgfplots}
\pgfplotsset{compat=1.18}
\usepgfplotslibrary{groupplots}
\pgfplotsset{every axis/.append style={font=\footnotesize, tick label style={font=\scriptsize},
  label style={font=\scriptsize}, grid style={gray!18}}}

\usepackage{subcaption}

\newcommand{\takeaway}[2]{%
  \par\bigskip\noindent
  \begin{tikzpicture}
    \node[draw=c0!55!black, line width=0.6pt, fill=c0!7, rounded corners=3pt,
          inner sep=6pt, text width=\dimexpr\columnwidth-15pt\relax, align=left]
      {\small\textbf{Takeaway (#1).}\hspace{2pt}#2};
  \end{tikzpicture}\par\smallskip
}

\begin{document}

\title{How Agentic Is Agentic Commerce? A Population-Scale Measurement of x402 Adoption and Authenticity}

\author{Shengchen Ling}
\affiliation{\institution{City University of Hong Kong}\country{Hong Kong SAR}}

\author{Yajin Zhou}
\affiliation{\institution{The Chinese University of Hong Kong}\country{Hong Kong SAR}}

\author{Lei Wu}
\affiliation{\institution{Zhejiang University}\country{China}}

\author{Cong Wang}
\affiliation{\institution{City University of Hong Kong}\country{Hong Kong SAR}}


\begin{abstract}
AI agents are said to be forming an economy in which they pay, on their own, for the data, APIs, and compute they consume. x402, which settles a stablecoin payment on-chain for each purchase, is the most widely deployed protocol for this, and its hundreds of millions of settlements are read as proof that the economy has arrived. We show the count cannot be read as adoption: it is the one metric an interested party can manufacture almost for free, since the facilitator sponsors the gas and nothing on-chain marks who controls a payment. We give the first population-scale measurement of x402 on Base, supplemented with a coarser Solana census. Identifying settlements from their on-chain event and resolving the true payer through the meta-transaction layer, we sort each by what its trace can prove via a payment graph. Over a 280-day window Base carries 136{,}708{,}672 settlements worth \$44{,}121{,}383.81, concentrated on every axis we measure (payer, recipient, and value Gini all above 0.98), yet 21.20\% are fictitious and 63.78\% internal settlement within a linked cluster. What is genuinely independent is bounded: it lies between the \$187{,}861.35 that demonstrably reaches a nameable service and the \$20{,}258{,}746.09 (45.92\% of value) not provably manufactured. Finally, we resolve the count's manufacturable component, a coherent operator-driven economy, star-shaped, machine-timed, and gas-subsidized. Settlement count measures manufacturability, not adoption.
\end{abstract}


\maketitle

\section{Introduction}
\label{sec:intro}

AI agents are increasingly cast as economic actors that will pay, on their own, for the data, APIs, and compute they consume. Agentic payment protocols such as x402 operationalize this. The x402 protocol revives the dormant HTTP 402 ``Payment Required'' status code and, each time an agent buys a resource, settles a stablecoin payment on-chain through an authorization. It is the most widely deployed agentic-payment protocol and the only one observable on-chain at population scale. The ecosystem now places ``agentic payment'' activity in the tens to hundreds of millions of settlements~\cite{x402scan,a16zaudit}, and these are read as evidence that an agent-driven economy has arrived. 

\textbf{The problem.} These counts are trusted as adoption, yet settlement count is the one metric an interested party can manufacture almost for free. Because the facilitator sponsors the gas, a producer can settle millions of times at near-zero marginal cost; and because nothing on-chain marks who controls a payment, a manufactured settlement is indistinguishable from genuine demand at the level of a single transfer, so a raw count conflates the two and no ground-truth label of genuineness exists to separate them. What an adoption claim actually needs is genuine demand, the activity that serves a need external to its own producer. The most a measurement can do is bound it and characterize its complement, the manufacturable component, and even that has lacked a rigorous, reproducible treatment.

\textbf{Research questions.} We ask three research questions, and answer each in its own section. 
\begin{enumerate}
    \item \textbf{RQ1} (\S\ref{sec:rq1}): how much x402 settles on-chain, how concentrated is it, and how much of the advertised supply actually materializes?
    \item \textbf{RQ2} (\S\ref{sec:rq2}): how much of that activity is genuine agent demand rather than operator-generated, and who generates it?
    \item \textbf{RQ3} (\S\ref{sec:rq3}): what is the manufactured economy that dominates the count, and why is settlement count therefore not a credible adoption metric?
\end{enumerate}

\textbf{Challenges.} Measuring x402 at population scale poses three challenges that off-the-shelf chain analysis cannot overcome. \emph{1) Identification.} No transaction is natively marked ``x402'': a settlement's only on-chain trace is a generic EIP-3009 authorization event, indistinguishable from any unrelated EIP-3009 transfer, and submitted by the facilitator's gas-sponsoring relayer rather than the paying agent. \emph{2) Authenticity.} A correctly identified settlement carries no signal of whether it answers demand external to its producer, and since genuine agent traffic is itself automated, periodic, and dust-priced, the signatures that betray manufacture in other markets do not discriminate here. \emph{3) Characterization.} The payment record itself directly reveals none of the machinery behind the settlements, so the manufactured component must be reconstructed from signals orthogonal to the payments: their funding, capital recycling, timing, and dynamics.

\textbf{This paper.} We present the first population-scale measurement of x402 on Base and supplemented with Solana. We identify settlements\footnote{A settlement is a transfer that emits the USDC \texttt{AuthorizationUsed} event.} by an event-defined rule intersected with a listed facilitator, recover the true payer through the meta-transaction layer, and report every quantity against an explicit base set and measure. Telling genuine demand from an operator's payments to internals rests on the one criterion that survives an agentic setting, whether value reaches a party outside the transactor's control or cluster. On a directed value-flow graph of settlement, funding, and sweep transfers, we resolve addresses into funding-linked clusters and sort each settlement by what its endpoints prove, calling it fictitious or internal settlement only when the trace establishes it and leaving the rest unattributed. The result bounds genuine demand. Finally, we characterize the operator economy that fills the count from signals orthogonal to the settlements, its structure, timing, capital, price, and dynamics.

\textbf{Findings.} At population scale, the surface metrics that pass for agentic-payment adoption are large but hollow (RQ1). Over a 280-day window Base carries 136{,}708{,}672 x402 settlements ($S$) worth \$44{,}121{,}383.81, with Solana adding 49{,}477{,}928 more under a coarser counting unit. Yet the activity is extremely concentrated on both chains and on every base set, with payer, recipient, and value Gini coefficients all above 0.98 and a single dominant facilitator on each chain. Its advertised supply barely materializes: 25{,}163 catalogued Base resources resolve to only 811 distinct recipients, of which 624 ever settle and just 249 ever earn at least \$10, while only 474 of the 910 advertised hosts (52.09\%) still return a live payment challenge. On authenticity (RQ2), that surface is overwhelmingly operator-internal: sorting every settlement by what its on-chain trace proves, 21.20\% of $S$ is fictitious (C1, self-payment or a provably closed loop) and a further 63.78\% is internal settlement inside a linked cluster (C2), leaving only 15.02\% unattributed (C3). The fictitious tier alone holds 54.08\% of the value, so at most 45.92\% (\$20{,}258{,}746.09) is not provably manufactured; within that ceiling only \$187{,}861.35 demonstrably reaches a nameable catalog service, the rest terminating at recipients we cannot name. The genuinely independent economy is therefore bounded between these two figures, two orders of magnitude apart. On RQ3, this operator-internal activity is a coherent operator-driven economy. Value flow forms a forest of operator-centered stars rather than a marketplace; it is machine-timed; and its dollars are recycled operator capital. Because the facilitator sponsors the gas, the entire 136{,}708{,}672-settlement count is reproducible for about \$355{,}583 in gas, paid by the facilitators. Its history is authored, the largest movements in the series being overwhelmingly the operators' own campaigns switching on and off, and its growth is intensive, a shrinking payer base each settling more. Settlement count is thus a Goodhart metric: gas-subsidized and reward-linked, it measures the incentive to transact, not the adoption it is taken to prove.

\textbf{Contributions.} We make the following contributions.
\begin{itemize}
\item \textbf{A measurement methodology and public artifact.} We give a reproducible pipeline that identifies x402 by settlement event and listed facilitator, resolves the true payer, imposes an explicit counting discipline, and classifies every settlement on a value-flow graph; upon publication we will release all data and implementation, so new windows, facilitators, and chains can be re-measured (\S\ref{sec:rq1},~\S\ref{sec:rq2}).
\item \textbf{The first population-scale measurement of x402 adoption and authenticity.} Over a 280-day window we census both deployment chains and establish that the headline count is overwhelmingly operator-internal, 84.98\% of Base's 136{,}708{,}672 settlements, within which at least 21.20\% in count or 54.08\% in value is manufactured; and that genuine third-party demand is well bounded (\S\ref{sec:rq2}).
\item \textbf{An anatomy of the manufactured economy.} From signals orthogonal to the settlements we show this operator-internal activity is a coherent operator-driven system, a forest of machine-timed stars circulating recycled capital, reproducible in its entirety for about \$355{,}583 in sponsored gas, so settlement count is a Goodhart metric that measures the incentive to transact rather than the adoption it is taken to prove (\S\ref{sec:rq3}).
\end{itemize}

\textbf{Organization.} \S\ref{sec:background} sets out the background on x402 and the agentic-payment setting and derives the value-closure criterion and three-tier scheme by which we tell manufactured from genuine activity. \S\ref{sec:rq1} answers RQ1, fixing the identification rule and measuring x402's scale, concentration, and supply. \S\ref{sec:rq2} answers RQ2, resolving operators and sorting every settlement into fictitious, internal, or unattributed on the value-flow graph to bound the genuinely independent economy. \S\ref{sec:rq3} answers RQ3, characterizing the manufactured economy along its structure, timing, capital, price, and dynamics. \S\ref{sec:limitations}-\ref{sec:related} are discussions and related work; and \S\ref{sec:conclusion} concludes. 

\section{Background}
\label{sec:background}

\subsection{x402 and Agentic Payments}
x402~\cite{x402spec} settles HTTP requests on-chain by layering HTTP 402 ``Payment Required'' over stablecoin transfer. A client, often an AI agent, requests a paid resource, receives a 402 with payment terms, and signs an EIP-3009 \texttt{transferWithAuthorization} authorization~\cite{eip3009}; a \emph{facilitator} then relays that signed authorization on-chain and pays gas on the client's behalf. The settlement asset is usually USDC and, on Base, the on-chain trace of a completed payment is the \texttt{AuthorizationUsed(address,bytes32)} event emitted by the USDC contract; on Solana there is no equivalent protocol event and a settlement is a USDC SPL \texttt{TransferChecked}. The meta-transaction layer entangles three roles that any correct count must keep distinct. The \emph{facilitator}, through its on-chain \emph{relayer} EOA, broadcasts and pays gas and appears as \texttt{transaction\_from}; the \emph{payer} is the authorization signer who consents to the transfer; and an \emph{operator} is the economic entity that may control many payer wallets and recipient hubs at once. 

\subsection{Defining Manufactured Activity}
\label{sec:genuine}
\subsubsection{Established Detection Criteria}
Separating manufactured from genuine on-chain activity rests on a toolkit built across wash-trading, Sybil, and invalid-traffic detection~\cite{sybil}, which we group into six criteria. (i) \emph{Self-dealing and beneficial-ownership change}: activity is inauthentic when one controller sits on both sides, either directly as a self-transfer or indirectly through a common funder that seeds the colluding accounts, and common funding is repeatedly reported as one of the strongest on-chain signals~\cite{trustasybil,fistful}. (ii) \emph{Closed loops with no net position change}: graph methods flag strongly connected components whose members transact almost only among themselves and whose balances are unchanged once all transfers net out~\cite{dexwash,nftwash}. (iii) \emph{Absence of economic substance}: volume that recycles the same capital without importing new capital or effecting a net transfer inflates counts without moving value~\cite{washtrade}. (iv) \emph{Value anomalies}: identical sizes, round numbers, and Benford-law deviations betray scripted amounts~\cite{washtrade}. (v) \emph{Coordinated lifecycle and timing}: nominal identities funded and activated together, in tight temporal clusters and at machine-regular intervals~\cite{trustasybil,financialbots}. (vi) \emph{Structural exclusivity}: colluding wallets concentrate on one venue and rarely touch independent participants~\cite{dexwash}. These criteria were calibrated on exchange order books, NFT marketplaces, and airdrop farming, settings in which the manipulator behaves visibly differently from a human trader or a diversified user.

\paragraph{Why agentic payments defeat most of them.} Agentic payments differ from those criteria. First, the legitimate actor is itself a bot: genuine x402 demand is produced by AI agents that are automated and high-frequency by construction~\cite{x402spec}, so automation, periodicity, and high volume largely cease to separate genuine from manufactured. Second, fixed-price micro-payments are the norm: an agent paying a metered API a fixed \$0.001 per call legitimately emits identical, round, dust-sized amounts. Third, custodial funding is ordinary: agent-wallet platforms custody their users' agents and top them up from a single treasury, so the funding signal is confounded; a legitimate custodial platform and a wash pool share the identical one-funder-to-many-payers shape. Fourth, settlements are mostly plain USDC transfers, carrying no price, spread, or order book, so the market-microstructure signals several detectors relies are absent.

\paragraph{What survives.} One criterion survives, because it turns on neither who transacts nor how, only on whether value reaches a party outside the transactor's control: economic substance. A facilitator only relays a payer's authorized transfer to the named seller and never returns funds, so no legitimate settlement path routes a hub's receipts back to the wallet that funded its payers. Funding a fleet of agent wallets from one treasury is itself ordinary and legitimate, the documented architecture of custodial agent-wallet providers~\cite{lnpay}, but a hub that sweeps its receipts back to that treasury has added a second, deliberate transfer that closes a loop. We therefore make value closure the primary test and measure it on a directed, layered value-flow graph whose edges are settlement, funding, and sweep transfers. We resolve each operator into a cluster and ask, of the value its settlements move, whether it terminates at parties outside the cluster or recirculates within.
Further, the timing criterion is not wholly neutralized. A genuine agent pays because some demand process drives it, which leaves variance in its cadence, whereas a wash script serving no demand can emit at a fixed interval with near-zero variance. Sustained zero-variance periodicity therefore retains evidential value, but it is not decisive, since a scheduled agent paying at exact intervals is also legitimate.

\subsubsection{Our Three-Tier Criteria}
We sort every settlement into three tiers by what its on-chain trace can prove~\footnote{Verdicts attach per settlement, and only to x402 settlements, the objects that $S$ counts; funding and sweep transactions are ordinary token transfers, never part of $S$, and enter only as evidence for resolving clusters. The verdict is by endpoint, not by payer: a settlement that reaches a payee outside its payer's cluster counts as unattributed (C3).}, asserting only the direction each tier supports and never that a settlement is genuine. \emph{Fictitious} (C1): the payer and payee are one wallet, or the settlement sits in a cluster that is provably closed, every transfer recirculating within the cluster and none crossing its boundary, so the value performs no on-chain economic work; this is the only tier we call outright manufactured. \emph{Internal settlement} (C2): the payer and payee resolve to one cluster linked by a shared bespoke seed funder or an improbable shared address-vanity pattern, so the settlement is internal to that linked cluster rather than a payment towards external parties, yet the cluster is not fully closed. We do not claim the cluster is one controlling entity; whether it is deliberate wash or a custodial platform's own internal ledger, its settlements cannot be read as independent adoption, exactly as a payment processor's internal ledger entries are not counted as merchant adoption. \emph{Unattributed} (C3): the payee lies outside the payer's cluster, or the payer is in no identified cluster; these settlements reach a party outside any resolved cluster, which may be independent or merely unresolved. Common funding identifies clusters but never by itself establishes single control, since custodial funding makes it necessary and not sufficient.

\section{RQ1: How Much x402 Settles On-Chain, and How Concentrated Is It?}
\label{sec:rq1}

\subsection{Identification}
\label{sec:idvalidity}

\subsubsection{Window and Roles}
We fix the study window to 2025-09-17 through 2026-06-23 (280 days).
Settlement is a meta-transaction, so its on-chain sender is the relayer, not the payer. We resolve roles before any clustering: the \emph{payer} $P$ is the authorization signer (equal to the USDC \texttt{Transfer.from} and to \texttt{AuthorizationUsed.authorizer}); the \emph{recipient} $R$ is \texttt{Transfer.to}; the \emph{relayer} $F$ is \texttt{transaction\_from}. Reading the signer from \texttt{AuthorizationUsed.authorizer} recovers the party that consented to the payment, and its agreement with \texttt{Transfer.from} serves as a per-settlement consistency check.

\subsubsection{Identification Rules}
An x402 settlement leaves the same on-chain trace as any other sponsored EIP-3009 transfer, and the only fact that makes it ``x402'', the HTTP 402 exchange, is off-chain. Identifying x402 on-chain is therefore a bounding problem. We define an x402 settlement as a USDC transfer that emits \texttt{AuthorizationUsed(address,bytes32)}, i.e., the EIP-3009 authorized-transfer marker, \emph{and} is relayed by a facilitator on the community-maintained list. We adopt this intersection because each half fails on its own, in opposite directions. First, counting every \texttt{AuthorizationUsed} event is a loose upper bound, since the token contract emits it for any EIP-3009 transfer. It admits the entire event universe of 140{,}274{,}969 settlements, gross \$772{,}090{,}109.10. Second, a broad relayer-address filter fails the other way, because the relayer EOAs also broadcast large non-x402 treasury and DeFi flow. Attributing every USDC transfer sent by the 5{,}587 addresses that ever relay a settlement would total \$42{,}348{,}540{,}765, 54.8$\times$ even the event-universe gross. Our rule is the intersection of the event with a curated facilitator allowlist, which tightens the universe to the 136{,}708{,}672 x402 settlements (\$44{,}121{,}383.81).

The ecosystem is permissionless and has no authoritative registry; therefore, we take the allowlist as the union of two independent curations of x402 facilitators. The first is Allium's on-chain index that labels facilitators~\cite{allium}; and the second is from the facilitator package of the ecosystem explorer, i.e., x402scan~\cite{x402scan}. The two agree closely, and totally we resolve 30 facilitators (details in Appendix~\ref{app:allowlist}). Because a single facilitator can operate several relayer EOAs, the union further resolves to 129 relayer EOAs, a strict subset of the 5{,}587 relayers that appear in the EIP-3009 event universe. 

This rule is specific to Base, whose USDC contract emits the \texttt{AuthorizationUsed} event. Solana exposes no equivalent event, so a Solana settlement is identified by the coarser criterion of a USDC SPL transfer whose fee-payer signer is a registered facilitator. On that facilitator set, the same two curations (Allium and x402scan) agree identically (15 of 15; Appendix~\ref{app:allowlist}). Therefore, the Solana count is only a confirmed-facilitator upper bound.

\subsubsection{Identification Result}
These results are the base dataset for the following sections. Every reported quantity names its base set and measure: settlements $S$, payers $P$, recipients $R$, and relayers $F$. On Base, the x402 census has $S=136{,}708{,}672$ settlements, $P=491{,}587$ payers, $R=116{,}781$ recipients, $F=129$ listed facilitator relayer EOAs; x402 gross value is \$44{,}121{,}383.81. On Solana, under the coarser criterion, the x402 census has $S=49{,}477{,}928$ settlements, $P=227{,}133$ payers, $R=23{,}743$ recipients, $F=23$ facilitator fee-payer signers; x402 gross value is \$9{,}032{,}706.12.

\subsection{Supply}
\label{sec:supply}
\begin{figure}[t]
      \centering
      \begin{tikzpicture}
      \begin{groupplot}[group style={group size=2 by 1, horizontal sep=22mm}, height=2.8cm, nodes near coords, nodes near coords align={horizontal}, point meta=explicit symbolic, every node near coord/.append style={font=\tiny, anchor=west, xshift=2pt}, tick label style={font=\scriptsize}, label style={font=\scriptsize}, title style={font=\scriptsize, yshift=-1.2mm}]
      \nextgroupplot[xbar, bar width=5pt, enlarge y limits=0.35, width=0.45\columnwidth, xmin=0, xmax=1050, symbolic y coords={earned $\geq$\$10, ever settled, distinct payTo}, ytick=data, xlabel={recipients}, title={(a) on-chain settlement funnel}]
        \addplot[fill=blue!55!black, draw=none] coordinates {(811,distinct payTo)[811] (624,ever settled)[624] (249,earned $\geq$\$10)[249]};
      \nextgroupplot[xbar, bar width=5pt, enlarge y limits=0.35, width=0.45\columnwidth, xmin=0, xmax=1150, symbolic y coords={live 402, any response, probed}, ytick=data, xlabel={hosts}, title={(b) off-chain HTTP liveness}]
        \addplot[fill=orange!80!black, draw=none] coordinates {(910,probed)[910] (805,any response)[805] (474,live 402)[474]};
      \end{groupplot}
      \end{tikzpicture}
      \caption{Base's x402 supply; each panel is a funnel over a single unit. (a) Recipients: the 25,163 advertised resources dedup to 811 distinct payTo, of which 624 (76.94\%) ever settle and 249 (39.90\% of those) earn at least \$10 of lifetime gross (median \$3.96). (b) Hosts: of the 910 distinct Base hosts, 88.46\% respond to a GET-only probe but only 52.09\% return a live 402 challenge. }
      \label{fig:funnel}
  \end{figure}
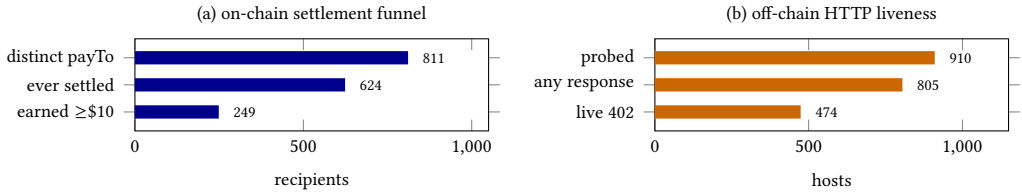

On-chain we observe only settlements; a resource that is advertised but never paid leaves no on-chain trace. The gap between advertised and used capacity is therefore visible only by matching the off-chain \textit{catalog} against the chain. To this end, we take the advertised surface from the CDP Bazaar~\cite{cdpx402docs}, the x402 protocol's resource-discovery catalog operated by Coinbase, which is the canonical channel a seller registers with to be found by agents. We discovered 25{,}411 resources over 37{,}338 payment accepts\footnote{This is the protocol's \texttt{accepts} entry; each represents a scheme/network/asset a resource will take payment in.}, of which 25{,}163 advertise a Base option, as of June 2026. We resolve each listing to its \emph{payTo} address, the on-chain recipient it names, and to its serving \emph{host}; the 25{,}163 Base resources resolve to 811 distinct payTo across 910 distinct hosts. We then match the payTo against the recipients $R$ of the on-chain settlement record (Fig.~\ref{fig:funnel}a).

The surface narrows sharply at every stage. Specifically, the 25{,}163 advertised Base resources collapse to only 811 distinct payTo recipients, because the catalog is mostly duplicated. Two payTo template factories (\texttt{lowpaymentfee.com}, 10{,}035 listings, and \texttt{orbisapi.com}, 9{,}461 listings) emit 77.48\% of it while earning roughly \$2{,}234 over their lifetime. Reaching the ledger is then common but earning is not: 624 of the 811 distinct recipients (76.94\%) ever settled at least one payment, yet only 249 of those (39.90\%) earned at least \$10 over their lifetime, at a median lifetime revenue of \$3.96 among the 624~\footnote{The \$10 floor is a presentation choice: of the 624, 423 earn at least \$1, 291 at least \$5, 249 at least \$10, 134 at least \$50, and 97 at least \$100, the same steep decay at every threshold.}. Because a single test transaction satisfies ``ever settled'', only the revenue threshold separates a serviced endpoint from one activated once.

Further, the off-chain surface has decayed in step. A single-round, GET-only ethical probe of the 910 distinct Base hosts (no payment performed, one advertised endpoint per host) finds 88.46\% (805) still return an HTTP response but only 52.09\% (474) a live HTTP~402 payment challenge, the rest returning no response at all (105) or a response without a challenge (331) (Fig.~\ref{fig:funnel}b). The figure is a point-in-time liveness estimate under this deliberately minimal probe design, not a stable live-service rate. Moreover, the ecosystem is also single-scheme and single-chain. The dynamic-pricing ``upto'' scheme is advertised but essentially dead. Of the 259 resources that carry it, 185 (71.43\%) come from a single address that has never settled a payment on Base (zero transactions, an unfunded EOA); The 11 addresses advertising it earn between \$285.88 (the 4 advertising only \texttt{upto}) and \$1{,}040.63 (all 11) over their lifetime, since \texttt{upto} and \texttt{exact} settlements to a mixed address are inseparable on-chain. Finally, the catalog's apparent multi-chain breadth is illusory. Of all 25{,}411 catalog entries across networks, 10{,}681 (42.03\%) advertise a Solana payment option, but 99.72\% (10{,}651) of them also advertise a Base option, so they are the same sellers offering one resource on two networks, not a Solana-native supply pool.

\subsection{Concentration}
The activity is extremely concentrated on every base set and on both chains. The payer, recipient, facilitator, and value distributions answer distinct questions (who sends, who receives, who relays, and who moves value), so the concentration is structural (Table~\ref{tab:concentration}). The evaluation metrics are explained here\footnote{The \emph{Gini coefficient} of a share distribution runs from 0 (all equal) to 1 (one entity holds everything). The \emph{Nakamoto coefficient} is the minimum number of top entities whose combined share first exceeds 50\%. The \emph{HHI} (Herfindahl-Hirschman Index) is the sum of squared percentage shares, from near 0 (fragmented) to 10{,}000 (a monopoly). Ginis are exact population values computed over the census, and the value Gini is over per-payer settled value. The bracketed 95\% payer-resampling bootstrap intervals (payer count [0.98309, 0.99036], recipient count [0.99530, 0.99884], payer value [0.97730, 0.99337]) are reported only to show insensitivity to any small set of addresses. }.
Consequently, the two chains reach comparable Gini extremes by different routes. On Base, the concentration is distributed across a few operators, the largest recipient taking 23.08\% of settlements (recipient Nakamoto 3) and the payer side spread more thinly (payer Nakamoto 112). On Solana it funnels into a single hub: the largest recipient alone takes 64.52\% (recipient Nakamoto 1), while the payer side is spread across many seeded wallets (payer Nakamoto 271, largest payer 1.93\%). The facilitator layer is one-operator on both chains (largest facilitator 63.54\% on Base, 67.72\% on Solana, facilitator Nakamoto 1 on both), but only overtly so on Solana: Base's dominant facilitator shards across 129 relayer EOAs, so at the EOA level it looks competitive (HHI 292) even though it is actually one entity load-balancing (\S\ref{sec:desybil}). 

  \begin{table}[t]
    \caption{Concentration of the x402 census on Base and Solana.}
    \label{tab:concentration}
    \small
    \setlength{\tabcolsep}{4.5pt}
    \begin{tabular*}{\columnwidth}{@{\extracolsep{\fill}}l ccc ccc ccc@{}}
      \toprule
      & \multicolumn{3}{c}{Gini} & \multicolumn{3}{c}{The largest (\% of $S$)} & \multicolumn{3}{c}{Nakamoto} \\
      \cmidrule(lr){2-4}\cmidrule(lr){5-7}\cmidrule(lr){8-10}
      & payer & recipient & value & payer & recipient & facilitator & payer & recipient & facilitator \\
      \midrule
      Base   & 0.98702 & 0.99808 & 0.98872 & 8.44\%  & 23.08\% & 63.54\% & 112 & 3 & 1 \\
      Solana & 0.98463 & 0.99526 & 0.98054 & 1.93\%  & 64.52\% & 67.72\% & 271 & 1 & 1 \\
      \bottomrule
    \end{tabular*}
  \end{table}
  
\takeaway{RQ1}{x402's principal metrics are large but hollow: a settlement count in the hundreds of millions, yet extreme concentration on every base set and on both chains, and an advertised catalog that overwhelmingly never settles or earns. What passes for population-scale adoption is, read straight off the ledger, thin and highly concentrated rather than broad.}

\section{RQ2: How Much Is Genuine Demand, and Who Generates It?}
\label{sec:rq2}

\S\ref{sec:rq1} found a large surface, but its raw settlement count cannot show how much of it is genuine demand, since an independent party paying for a service and an operator paying among addresses it controls leave identical counts. This section tells the two apart, sorting every settlement by what its on-chain trace can prove. 

\subsection{Resolving Operators}
\label{sec:desybil}
A service operator can own many addresses, so before activity can be grouped we resolve their wallets into clusters. Two conservative rules link wallets into one cluster when 1) they share an exact funder (often at a fixed amount), or 2) a vanity-suffix address batch. Both key on an on-chain trace that is hard to forge by accident. 
Resolved this way, the operators are few but each is large. Named operators account for \textbf{67.76\%} of $S$ (92,640,016 settlements), the three largest being lnpay (23.08\%), t54 (21.74\%), and the 2025-11 self-payment activity (20.13\%). Each name is read off the catalog itself, where the operator advertises its recipient hub under its own domain\footnote{The lnpay hub is under \texttt{pay.lnpay.ai} and \texttt{market.lnpay.ai}, t54's under \texttt{x402-secure-api.t54.ai}.}. Naming and attribution are separate steps: which wallets form a cluster is fixed on-chain by the merge rules above, whereas the name is only the public handle its hub advertises in the catalog. 

Resolving wallets to their operators is what collapses the apparent breadth of \S\ref{sec:rq1}. Counted as raw wallets, the payer side looks like a broad network, with a Nakamoto coefficient of 112 (Base); while counted as the operators behind, the coefficient is 3. Value tells the same story, at a value-level Nakamoto of 3.
The relayer layer carries the same illusion. The 129 x402 relayer EOAs give a low HHI (292), which would ordinarily read as a competitive market of many independent relayers. But a genuine market does not manufacture equal-sized competitors, and here the relayers ranked 4th through 12th each carry between 3.24\% and 3.26\% of settlements, nine shares equal to within roughly two hundredths of a percentage point. Independent parties do not land on identical shares by chance; near-perfect equality across a run of ranks is the fingerprint of one operator's round-robin load balancer spreading its own traffic evenly over its EOAs.

\subsection{Graph-Based Three-Tier Classification}
\label{sec:method-graph}
We operationalize the three-tier criteria of \S\ref{sec:genuine} on the whole population with a graph method: it builds a layered value-flow graph, resolves operators into clusters, and sorts every settlement into fictitious (C1), internal settlement (C2), or unattributed (C3) by what its on-chain trace proves.

\subsubsection{The Value-Flow Graph and Clustering}
We build the directed value-flow graph: nodes are addresses, and edges are USDC transfers in three roles, x402 settlements (payer to hub), funding (a wallet's inbound capital), and sweeps (a hub's outbound transfers). Around the recipient hubs that concentrate settlement mass, we resolve funding-linked clusters: for every payer of a hub we read its funder on-chain, and a payer base tracing to one bespoke funder forms a cluster spanning funder, fleet, and hub\footnote{These clusters are the operators resolved in \S\ref{sec:desybil}~\cite{fistful}, with exchange, router, and relayer hubs held out as opaque boundaries so a high-degree shared address cannot falsely merge unrelated operators. We confirm the fund-cycling structure independently, by strongly-connected-component and cycle enumeration on the settlement layer~\cite{dexwash}.}. We then sort every settlement by what its endpoints prove. First, two patterns are fictitious (C1). One is self-payment, where payer and payee share the same address; the other is a cluster that is provably closed, its fleet settling only to its own hub and the hub returning that value to the cluster with no external outflow. Second, a settlement inside a funding-linked cluster that is not provably closed is internal settlement (C2): payer and payee fall in the same cluster, so it is not an independent-party payment, though we do not adjudicate whether it sits atop genuine off-chain demand. Third, every remaining settlement reaches a party outside any resolved cluster and is unattributed (C3). Where a hub is a real service drawing common payers alongside a seeded fleet, only the seeded fleet is placed in C2 while the external payers' settlements stay in C3 (Fig.~\ref{fig:operators}).

\subsubsection{Fictitious Activity (C1)}
\begin{figure}[t]
  \centering
  \includegraphics[width=0.9\columnwidth]{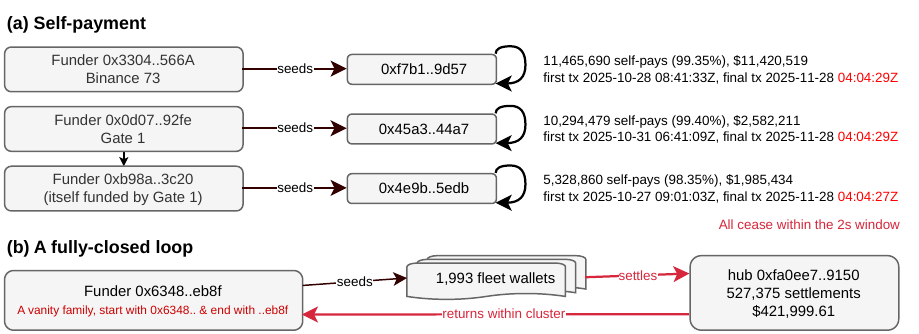}
  \caption{Two patterns of the fictitious tier (C1), where value performs no on-chain economic work. (a) Self-payment: wallets that pay themselves. (b) A fully-closed loop: a seed funder's fleet settles back only to the operator's hub, so no value crosses the cluster boundary. Both differ from the near-closed operators of Fig.~\ref{fig:operators}, which leak a little value and count as internal settlement (C2).}
  \label{fig:trace}
\end{figure}

The fictitious tier is the activity that provably performs no on-chain economic work; the two typical patterns are demonstrated in Fig.~\ref{fig:trace}. First, self-payment, where payer and payee are the same address, has no distinct counterparty and moves no value out of one wallet; this alone is 20.13\% of $S$. Beyond it, a cluster is also fictitious when it is provably closed: its fleet settles solely to the operator's own hub and the hub's receipts never leave the cluster, so no value reaches any party outside it. Four clusters meet this strictly, with exactly zero external settlements and exactly zero external value. \textbf{In total, the fictitious tier is 21.20\% of $S$}, provable from on-chain traces alone. We deliberately do not promote the many \emph{near}-closed operators that leak even a cent. For instance, lnpay drives 31{,}551{,}433 settlements into its hub and returns 98.34\% of the \$415{,}765 inflow to its funder; the residue does leave the cluster, so its settlements are classified as internal settlement (C2). This conservatism keeps C1 an unarguable lower bound.

Specifically, the self-payment is almost entirely a 2025-11 campaign of three wallets that settle almost only to themselves, 27{,}089{,}029 of the 27{,}514{,}640 self-settlements (98.45\%). First-funded from two exchange wallets (Binance and Gate), all three nonetheless cease within the same two-second window (last settlement 2025-11-28 04:04:27--29Z). The four fully-closed clusters run the same seed-and-return mesh: a funder seeds a fleet that settles only to the operator's own hub, and the hub sweeps its entire receipts back within the cluster. Tracing each hub's off-rail USDC by RPC confirms the closure to the dollar. The four hubs move \$779{,}488.96 between them with external outflow of exactly \$0.00, the largest returning its whole \$422{,}064.04 to its own funder. 

\subsubsection{Internal Settlement (C2)}
\label{sec:census}

The graph further decides C2 from on-chain traces: \textbf{the population is 63.78\% internal settlement (C2).} Each cluster we define an \emph{inflation factor}, its total settlements divided by those that terminate outside the cluster. A fully closed cluster (C1) has an infinite factor, and a cluster whose members mostly pay parties outside the cluster has a factor near one. Across the internal-settlement clusters these span orders of magnitude, with the highest above 7{,}801$\times$. The count and value factors need not agree, diverging where the internal circulation is dust but the little that leaves is not.

\begin{figure}[t]
  \centering
  \includegraphics[width=\columnwidth]{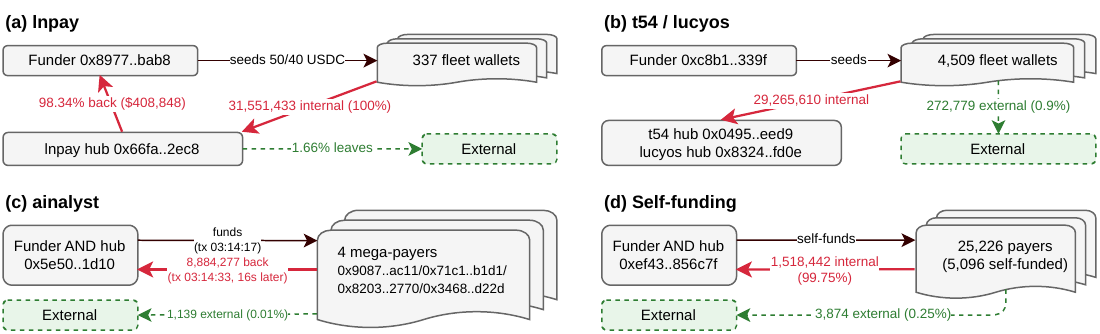}
  \caption{The four largest internal-settlement (C2) clusters. Red edges are internal circulation (fleet into the hub, hub back to the funder); green dashed edges are the value that reaches an external party. }
  \label{fig:operators}
\end{figure}

Concretely (Fig.~\ref{fig:operators}), a funder puts in 10{,}000.00 USDC and seeds lnpay's 337 fleet wallets in identical 50.00 or 40.00 USDC amounts, then the fleet wallets settle 31{,}551{,}433 transactions into the lnpay hub and none to an external seller, and later 98.34\% of the hub's \$415{,}765 inflow (\$408{,}848) returns straight to that funder. Because the residue does leave the cluster, it is C2. The fleet's payments are themselves funded by \$1{,}261{,}430 of x402 receipts, 99.98\% of which arrive from other wallets in the same fleet (\$314 from outside), so the money reaching the hub is the operator's own capital recirculating through a closed mesh. The t54 and lucyos operator drives 29{,}538{,}389 settlements with only 272{,}779 (0.9\%) reaching outside, 70.33\% of its payers first settling on one synchronized day; and ainalyst's four mega-payers drive 8{,}884{,}277 settlements against 1{,}139 external, the hub repaid 16 seconds after it funds a payer. 

\subsubsection{Unattributed (C3)}

\textbf{The remaining 15.02\% of $S$ is unattributed (C3)}: 20{,}529{,}628 settlements that reach a party outside any resolved cluster, carrying \$12{,}835{,}401.84 (29.09\% of value). This tail is where the graph stops; instead, we propose to use a residual model, which judges the C3 tail from an orthogonal angle, i.e., the payer's own behavior, asking whether an unclustered payer nonetheless behaves like a manufactured one. It asserts only the ``manufactured'' direction, never that a payer is positively genuine.
Trained on the human silver set using behavioral features alone, three model families agree on the C3 tail: applied to its 457{,}400 payers, a majority vote flags 87.82\% as ``manufactured-behaving'' (95\% CI [87.73\%, 87.92\%]). It is notable that we read this as an estimate of an otherwise unresolvable residual, not a validation of genuineness. Genuine agentic payments and manufactured traffic share multiple signatures (\S\ref{sec:genuine}), so annotators, and any model trained on their labels, can carry a stereotype bias. The modelling details are in Appendix~\ref{app:goldset}.

\subsection{Bounding the Genuine Economy}
\label{sec:residual}
Genuine economic activity is bounded above by C2 and C3 together, everything the chain does \textit{not} prove fictitious; only C1 is certainly manufactured, while C2 is internal settlement whose on-chain dust could in principle sit atop a custodial platform's real off-chain demand, so we cannot exclude it. Counting only C3, the value that actually reaches a party outside any cluster, the genuine independent economy is at most \$12{,}835{,}401.84 (29.09\% of the total); granting the benefit of the doubt to C2's \$7{,}423{,}344.25, it is at most C2 and C3 together, \$20{,}258{,}746.09 (45.92\%). Either way it is less than half of the \$44{,}121{,}383.81 headline.

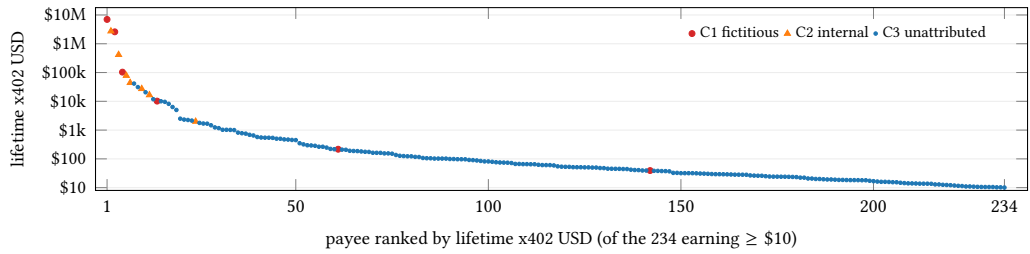
\begin{figure}[t]
  \centering
  \begin{tikzpicture}
  \begin{axis}[width=\columnwidth, height=4cm,
    xlabel={payee ranked by lifetime x402 USD (of the 234 earning $\geq\$10$)}, ylabel={lifetime x402 USD},
    ymode=log, xmin=-2, xmax=240, ymin=8, ymax=1.8e7, log basis y={10},
    ytick={10,100,1000,10000,100000,1000000,10000000},
    yticklabels={\$10,\$100,\$1k,\$10k,\$100k,\$1M,\$10M},
    xtick={1,50,100,150,200,234},
    tick label style={font=\scriptsize}, label style={font=\scriptsize},
    legend style={font=\tiny, at={(0.97,0.96)}, anchor=north east, draw=none, fill=none, legend columns=3},
    ymajorgrids, grid style={gray!14}]
    \addplot[c1, mark=*, mark size=1.1pt, only marks] coordinates {(1,6996030.82) (3,2596764.10) (5,103214.35) (14,10161.30) (61,216.18) (142,39.29)};
    \addplot[c4, mark=triangle*, mark size=1.3pt, only marks] coordinates {(2,2737208.51) (4,415760.71) (6,78070.98) (7,44092.76) (10,27196.99) (12,16544.72) (24,1988.02)};
    \addplot[c0, mark=*, mark size=0.65pt, only marks] coordinates {(8,41321.55) (9,31220.84) (11,20592.76) (13,11832.09) (15,10039.00) (16,9576.51) (17,8166.95) (18,6340.09) (19,5009.80) (20,2495.64) (21,2316.02) (22,2245.44) (23,2139.60) (25,1776.90) (26,1690.90) (27,1671.04) (28,1470.15) (29,1239.66) (30,1168.67) (31,1024.81) (32,1024.33) (33,1015.17) (34,1000.06) (35,817.24) (36,778.07) (37,752.62) (38,685.10) (39,649.98) (40,574.94) (41,552.68) (42,544.38) (43,542.68) (44,536.54) (45,502.81) (46,500.26) (47,474.94) (48,468.18) (49,454.71) (50,448.98) (51,349.66) (52,319.23) (53,297.43) (54,292.56) (55,284.67) (56,264.15) (57,263.92) (58,246.54) (59,222.01) (60,216.56) (62,209.32) (63,206.84) (64,190.82) (65,187.97) (66,186.36) (67,180.98) (68,175.86) (69,174.34) (70,162.70) (71,162.21) (72,161.68) (73,154.99) (74,154.97) (75,151.34) (76,136.38) (77,127.97) (78,125.13) (79,122.86) (80,122.52) (81,117.27) (82,115.97) (83,107.41) (84,105.30) (85,104.69) (86,102.48) (87,102.40) (88,102.37) (89,102.18) (90,98.99) (91,98.66) (92,97.23) (93,96.90) (94,96.57) (95,91.81) (96,90.52) (97,88.10) (98,84.73) (99,81.62) (100,81.42) (101,79.01) (102,76.37) (103,74.90) (104,74.24) (105,73.60) (106,71.74) (107,66.70) (108,66.61) (109,65.68) (110,65.65) (111,64.99) (112,64.98) (113,62.34) (114,60.88) (115,60.77) (116,60.67) (117,59.16) (118,54.99) (119,53.56) (120,52.91) (121,52.74) (122,51.49) (123,51.23) (124,51.19) (125,50.69) (126,50.63) (127,49.88) (128,49.85) (129,48.05) (130,47.72) (131,45.61) (132,45.39) (133,45.12) (134,44.99) (135,44.38) (136,44.29) (137,42.06) (138,40.99) (139,40.69) (140,39.50) (141,39.32) (143,38.35) (144,38.21) (145,37.68) (146,37.55) (147,36.84) (148,32.59) (149,32.59) (150,31.84) (151,31.84) (152,31.83) (153,31.82) (154,31.50) (155,30.94) (156,30.69) (157,30.24) (158,29.71) (159,29.40) (160,29.36) (161,29.21) (162,29.00) (163,28.67) (164,28.30) (165,28.24) (166,28.17) (167,28.00) (168,26.64) (169,26.22) (170,25.95) (171,25.88) (172,25.00) (173,24.37) (174,24.14) (175,24.12) (176,23.99) (177,23.94) (178,23.77) (179,23.56) (180,23.26) (181,22.31) (182,22.22) (183,20.80) (184,20.70) (185,20.04) (186,20.00) (187,19.54) (188,19.25) (189,19.23) (190,18.93) (191,18.57) (192,18.56) (193,18.37) (194,18.27) (195,18.27) (196,18.22) (197,18.08) (198,18.06) (199,17.33) (200,16.80) (201,16.30) (202,15.93) (203,15.92) (204,15.76) (205,15.47) (206,15.40) (207,14.78) (208,14.40) (209,14.11) (210,14.04) (211,13.95) (212,13.78) (213,13.75) (214,13.74) (215,13.60) (216,12.95) (217,12.94) (218,12.55) (219,12.37) (220,12.17) (221,11.85) (222,11.56) (223,11.36) (224,11.00) (225,11.00) (226,10.83) (227,10.67) (228,10.57) (229,10.52) (230,10.50) (231,10.50) (232,10.26) (233,10.25) (234,10.09)};
    \legend{C1 fictitious, C2 internal, C3 unattributed}
  \end{axis}
  \end{tikzpicture}
  \caption{The 234 catalog sellers that took in at least \$10 of lifetime x402 value\protect\footref{fn:ge10scope}, ranked by x402 USD (log scale) and colored by the tier their inbound settlements fall in. The largest earners are C1 and C2 cluster hubs, while C3 payees form the low tail.}
  \label{fig:genuine-ranksize}
\end{figure}

Within this ceiling the money reaches very few sellers. Of the catalog sellers that ever settled, only 234 took in at least \$10 of lifetime x402 value\footnote{Fewer than the 249 in the supply funnel (\S\ref{sec:supply}), which counts each advertised seller's total lifetime revenue; here we restrict to x402 settlement value, so sellers whose \$10 is largely non-x402 fall below the floor.\label{fn:ge10scope}}; below that floor a payee's few dollars are indistinguishable from testing traffics. Ranked by x402 earnings and colored by the tier its inbound settlements fall in, the distribution collapses more than five orders of magnitude, and the largest earners are C1 and C2 cluster hubs (Fig.~\ref{fig:genuine-ranksize}). On the other hand, considering only the C3 low tail, the result is even more stark. Of C3's \$12{,}835{,}401.84, only \$187{,}861.35 reaches an advertised catalog service, the rest terminating at recipients outside the catalog (unnamed but possibly legitimate private services, or ordinary wallets). We read that \$187{,}861.35 as the unattributed value that demonstrably reaches a nameable catalog service, the strongest affirmative trace of independent service revenue the chain offers. That our rules resolve no link between a payer and a catalog payee does not prove the two are unrelated, and conversely the off-catalog remainder may contain private services we simply cannot name. The genuinely independent economy is therefore bounded between \$187{,}861.35 and the \$20{,}258{,}746.09 ceiling, a two-order-of-magnitude range within which on-chain data cannot place it. 

\takeaway{RQ2}{Sorting every settlement by what its on-chain trace can prove, 84.98\% of the count is operators settling among their own addresses: 21.20\% provably fictitious (C1) and 63.78\% internal settlement within a linked cluster (C2), the remaining 15.02\% (C3) reaching a party outside any cluster. The genuinely independent economy is only bounded, between the \$187{,}861.35 that demonstrably reaches a nameable service and the \$20{,}258{,}746.09 ceiling, and cannot be pinned down from the chain; what the population-scale count resolves is its manufacturable component, not independent adoption.}

\section{RQ3: What Is the Manufactured Economy?}
\label{sec:rq3}

\S\ref{sec:rq2} bounded the genuine economy and found it small; what fills the settlement count instead is the overwhelmingly fictitious and operator-internal activity. This section asks what kind of system produces those activities, and traces it along five links. We begin with its structure (\S\ref{sec:recovery}), showing that value flow organizes into a forest of operator stars. We then examine its behavioral signature (\S\ref{sec:signature}), where the activity is machine-regular and demand-independent. Next, we interpret its value flow (\S\ref{sec:value-calibers}), finding that the observed settlement dollars largely represent recycled operator capital. We then explain why this activity scales so easily (\S\ref{sec:manipulability}): facilitator-sponsored gas reduces the marginal cost of additional settlements to nearly zero. Finally, we examine how the ecosystem evolves (\S\ref{sec:dynamics}), showing that growth is driven by particular campaigns. 

\subsection{Structure: A Star Forest}
\label{sec:recovery}
The settlement flow resolves into a forest of operator-centered stars, each operator at the hub of a payer fleet that settles almost only to that hub. The stars barely interpenetrate, sharing only a thin crust of low-volume payers; the any-to-any trading of a marketplace is absent.

\subsubsection{Graph Construction}
\begin{figure}[t]
  \centering
\includegraphics[width=\columnwidth]{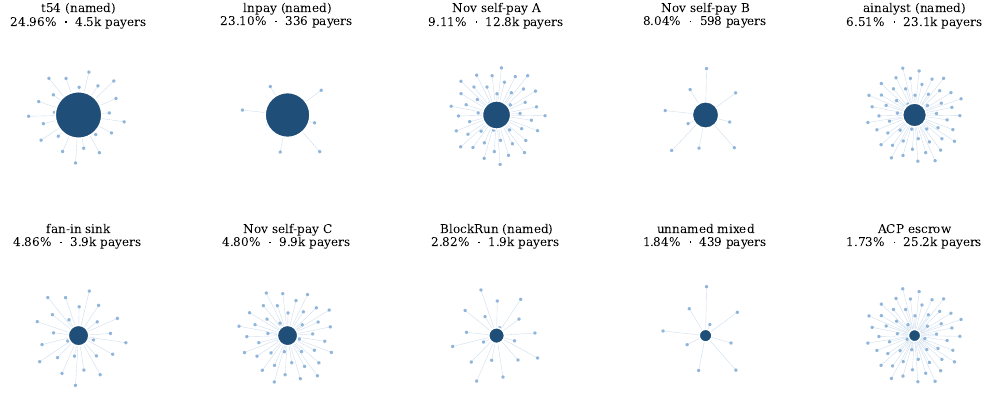}
 \caption{The 10 leading communities from label propagation, each an operator-centered star. In every panel the disc radius is fixed; the centre-marker area encodes settlement share and the number of leaf dots encodes payer count (square-root-scaled). }
  \label{fig:communities}
\end{figure}

We analyze the settlement layer of the value-flow graph of \S\ref{sec:method-graph}; where \S\ref{sec:desybil} resolved operators from its funding and sweep layers, here we read it by its connectivity alone. It is a directed, weighted graph whose nodes are addresses and whose edge $u \to v$ carries weight equal to the settlement count from payer $u$ to recipient $v$, with relayers excluded by construction\footnote{A relayer touches every settlement it forwards; if included, it would bridge disjoint operators into one artificial component and pose as the center of economic activity.}. The graph has 536,690 nodes and 718,668 edges; its giant weakly-connected component (409,360 nodes, 76.27\%) carries 99.2671\% of settlements, so we analyze that component and drop only a long tail of isolated dyads.
Unsupervised community detection makes the star forest explicit. By weighted asynchronous label propagation, in which each node adopts the settlement-weighted majority label of its neighbors, the graph partitions into operator-centered stars~\footnote{The partition is not seed-specific: across 10 label-propagation seeds, the t54 community's membership Jaccard against the reference run is 0.9976--0.9995 (median 0.9983) and lnpay's is 0.9227--0.9624 (median 0.9444), so every seed recovers the same operator stars. The Jaccard index is the number of wallets two runs place in the same community divided by the number placed in either, so 1 denotes identical membership and lower values a smaller overlap.}, with 10 leading communities shown in Fig.~\ref{fig:communities}. The largest by settlement volume are t54 (24.96\%) and lnpay (23.10\%); and the top-20 stars together carry 93.03\% of all settlements.

\subsubsection{A Forest of Operator Stars}
Since \S\ref{sec:rq2} established that most of this activity is operator-internal, we now describe the \emph{shape} its value flow takes. That shape is a forest of isolated operator stars rather than a machine-to-machine marketplace in which autonomous agents transact across many counterparties. We characterize it along six structural properties (Table~\ref{tab:marketplace}), each isolating a different feature of a mutual-trading market: whether value flows both ways (reciprocity), whether the graph stays a clean payers-to-hubs split (bipartivity), whether value is routed through intermediaries (a strongly-connected core), whether the busiest nodes are themselves densely interconnected (the dense $k$-core), and whether sellers share customers and stay connected (co-payer overlap and component count). On every one the observed graph sits at the star-forest extreme. It is notable that we only read these as a structural description of this economy, not as a claim of manufacture or genuine demand: a legitimate buyer-to-merchant or metered-API billing network can itself be low-reciprocity, sink-heavy, and near-bipartite. Graph-structural probes of this kind are a standard forensic handle on coordinated on-chain trading~\cite{nftsec,dexwash}.

\begin{table*}[t]
 \caption{Six structural probes of the payment graph. Each summarizes a property that distinguishes a mutual-trading marketplace from a forest of operator stars; on every one the observed graph sits at the star-forest extreme. The probes describe the flow's shape and are not by themselves a test of manufacture.}
  \label{tab:marketplace}
  \footnotesize
  \begin{tabular}{@{}lll@{}}
    \toprule
    Structural probe & A mutual-trading marketplace & Observed \\
    \midrule
    Reciprocity $\rho$~\cite{garlaschelli}       & value flows both ways  & $0.024930$, near zero \\
    Edge bipartivity~\cite{estrada}              & sellers trade sellers        & $0.9642$ of the maximum, nearly bipartite \\
    Intermediation~\cite{tarjan}  & value routed through chains & near-empty core; $20.40\%$ sink-only \\
    Dense $k$-core~\cite{seidman}  & the busiest nodes form the dense core              & disjoint \\
    Co-payer Jaccard overlap~\cite{jaccard,newman2001}  & sellers share buyers & $0.0036$ (hardly share payers) \\
    Co-payer components~\cite{newman2001}  & one connected component & $6$--$8$ disconnected components \\
    \bottomrule
  \end{tabular}
\end{table*}

First, a marketplace has value flowing both ways and passing through intermediaries, while here it does neither; reciprocity, the tendency of an edge $A\to B$ to be met by a return edge $B\to A$, is near zero (Garlaschelli-Loffredo~\cite{garlaschelli} $\rho = 0.024930$). Second, edge bipartivity~\cite{estrada}, how close the graph is to a clean payers-to-hubs split with no within-side edges, is 0.9642 of the maximum. Third, the strongly-connected core~\cite{tarjan}, which would hold any $A\to B\to C$ routing, is near-empty, and 20.40\% of settlements terminate at recipients that never pay onward: value moves one way, payer to hub, and stops.
Fourth, in a marketplace the highest-volume nodes form a dense, mutually-trading core. Here the densest region of the graph, its 58-core~\cite{seidman} (the largest set in which every node links to at least 58 others), holds only 94 nodes and is \emph{disjoint} from the high-volume operators, whose hubs sit at low core numbers. Fifth, the hubs' payer bases barely overlap: the top-40 hubs' payer sets have a mean pairwise Jaccard overlap~\cite{jaccard} (shared fraction of the two sets' union) of 0.0036, with the median pair sharing no payers at all. And sixth, they split into 6 to 8 disconnected components~\cite{newman2001} even when every bridge payer is retained. 

The same star-forest structure also recurs on Solana. Solana exposes no \texttt{AuthorizationUsed} analog and no full edge dump, so the edge-native community detection cannot run; we instead read the structure from server-side aggregates and an off-graph funder check over the confirmed-facilitator population. A single funder first-funds 56.59\% of all 227,133 payers (funder-share HHI 0.4043 across 5,493 distinct funders) under a fixed dust-ticket menu (0.05 USDC for 35.37\% of settlements, the top five amounts 92.94\%), similar to the Base fleets. On the list-complete 49{,}477{,}928-settlement set, 95.44\% of payers send at least 90\% of their settlements to a single payee, and those single-counterparty payers carry 91.68\% of settlements. The top payee alone absorbs 64.52\% (with top3 77.81\% and top10 93.11\%).

\subsection{The Signature: Machine Timing versus Human Demand}
\label{sec:signature}
The second link is how the activity behaves in time. We read two features measuring \emph{when} and \emph{how long} payers transact from settlement timestamps, over three populations set: the payers of provably-manufactured C1, the payers of operator clusters C2, and the payers of services that appear in the advertised catalog but resolve to no operator. The time-of-day clock and the survival-curve shape effectively separate advertised payers from the operator side.

\subsubsection{Time of Day, the Clock} 
\label{sec:clock_amplitude}
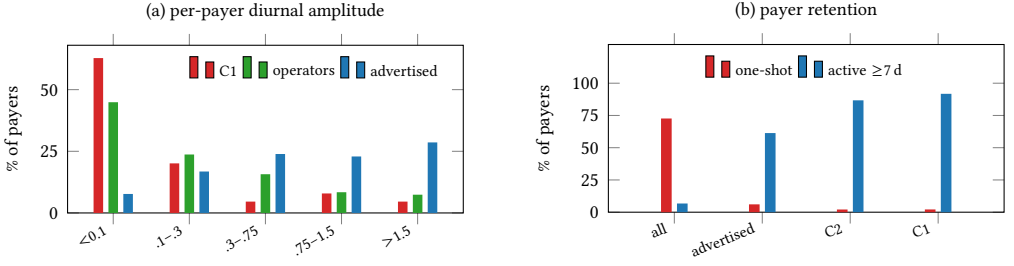
\begin{figure}[t]
  \centering
  \begin{minipage}{0.49\columnwidth}\centering
  \begin{tikzpicture}
  \begin{axis}[ybar, width=\columnwidth, height=3.8cm, ymin=0, ymax=68,
    xmin=0.4, xmax=5.6, xtick={1,2,3,4,5},
    xticklabels={{$<$0.1},{.1--.3},{.3--.75},{.75--1.5},{$>$1.5}},
    x tick label style={font=\tiny, rotate=30, anchor=east},
    bar width=3.6pt, ylabel={\% of payers}, ylabel style={font=\scriptsize}, ytick={0,25,50},
    tick label style={font=\scriptsize}, title={(a) per-payer diurnal amplitude}, title style={font=\scriptsize},
    legend style={font=\tiny, at={(0.97,0.97)}, anchor=north east, legend columns=3, draw=none, fill=none}]
    \addplot[fill=c1, draw=none] coordinates {(1,62.8) (2,20.1) (3,4.6) (4,7.9) (5,4.6)};
    \addplot[fill=c2, draw=none] coordinates {(1,44.9) (2,23.7) (3,15.7) (4,8.4) (5,7.4)};
    \addplot[fill=c0, draw=none] coordinates {(1,7.7) (2,16.8) (3,23.9) (4,22.9) (5,28.6)};
    \legend{C1, operators, advertised}
  \end{axis}
  \end{tikzpicture}
  \end{minipage}\hfill
  \begin{minipage}{0.49\columnwidth}\centering
  \begin{tikzpicture}
  \begin{axis}[ybar, width=\columnwidth, height=3.8cm, ymin=0, ymax=130,
    symbolic x coords={all, advertised, C2, C1}, xtick=data, bar width=4pt, enlarge x limits=0.25,
    ylabel={\% of payers}, ylabel style={font=\scriptsize}, ytick={0,25,50,75,100},
    tick label style={font=\scriptsize}, x tick label style={font=\tiny, rotate=22, anchor=east},
    title={(b) payer retention}, title style={font=\scriptsize},
    legend style={font=\tiny, at={(0.5,0.97)}, anchor=north, legend columns=2, draw=none}]
    \addplot[fill=c1, draw=none] coordinates {(all,72.64) (advertised,6.08) (C2,2.09) (C1,2.13)};
    \addplot[fill=c0, draw=none] coordinates {(all,6.74) (advertised,61.34) (C2,86.72) (C1,91.77)};
    \legend{one-shot, active $\geq$7\,d}
  \end{axis}
  \end{tikzpicture}
  \end{minipage}
 \caption{Two behavioral separations from settlement timestamps. (a) Per-payer hour-of-day amplitude (\S\ref{sec:clock_amplitude}): advertised-service payers are diurnal (median 0.79), while the provably-manufactured C1 payers and the C2 operator clusters are near-flat (0.08 and 0.12). (b) Payer retention (\S\ref{sec:retention}): all payers are mostly one-shot, C1 and C2 are provisioned and persistent, advertised services sit in between.}
  \label{fig:behavioral}
\end{figure}

A script does not sleep, so a machine-run emitter transacts at a flat rate around the clock, while a human-metered payer shows a day-night rhythm. We summarize each payer's daily rhythm by its \emph{hour-of-day amplitude}, the fundamental-harmonic amplitude of its settlement rate over the 24 hours of the day as a fraction of the mean (0 is flat around the clock, higher is concentrated in a daily window), as detailed in Fig.~\ref{fig:behavioral}a. The measure depends only on how concentrated a payer's activity is, irrespective of which hours it peaks in, so it is unchanged by the payer's time zone. Advertised-service payers carry a median amplitude of 0.79, against 0.12 for the operator clusters and 0.08 for the provably-manufactured C1 payers, each over its $\geq$50-settlement payers (the two largest operator hubs are near-flat, amplitude 0.010 for lnpay and 0.032 for t54). Against the cleanest manufacturing baseline, the C1 payers, the amplitude separates the advertised payers at AUC\footnote{Area under the ROC curve~\cite{aucroc}, 0.5 is chance and 1 is perfect separation.} 0.85. Against the operator clusters the AUC is lower, 0.78, because the operator population has a diurnal tail (7.4\% of operator payers score above 1.5). The diurnal tail also illustrates that an operator cluster is a structural grouping rather than a manufacturing verdict; some operator traffic may truly sit atop genuine human-metered demand. A similar flatness also holds on Solana, where the amplitude is 0.107 and peak-to-mean is 1.13, near the Base operator side and far from the advertised-payer rhythm. We adopt the payer as the unit because a service's aggregate clock is a superposition of its users' clocks, therefore a service whose users span time zones therefore reads flat in aggregate even while each user stays diurnal. 

\subsubsection{Lifespan Shape, the Survival Curve} 
\label{sec:retention}
A second axis is how long a payer persists, and the naive ``one-shot wallet is a bot'' heuristic fails outright: the C1 and C2 operator payers are the \emph{least} one-shot populations in the census (2.13\% and 2.09\% active on a single day, against 72.64\% network-wide and 6.08\% for advertised-service payers), as detailed in Fig.~\ref{fig:behavioral}b. The discriminating quantity is the shape of the survival curve. Specifically, of the C1 payers active over one week, almost all survive past four (a four-over-one-week ratio of 0.94, median active span 220 days), and the C2 operator clusters hold a ratio of 0.80 (median span 48 days). Advertised services instead decay more gradually (ratio 0.59, median span 16 days), the signature of users arriving and churning continuously. 

\subsection{Purchasing Power or Recycled Capital?}
\label{sec:value-calibers}

The third link is the capital. \S\ref{sec:rq2} showed the settlement \emph{count} is overwhelmingly operator-internal; are the \emph{dollars} any more real? Within x402 they are not. The fictitious tier alone carries 54.08\% of x402 value (\$23{,}862{,}637.72), and the genuine ceiling, everything the chain does not prove fictitious (the internal-settlement plus unattributed tiers), is \$20{,}258{,}746.09 (45.92\%). Inside that ceiling, the value that demonstrably reaches a nameable service is \$187{,}861.35, a floor two orders of magnitude below the ceiling itself. At least half the volume is provably not purchasing power, and where the genuine remainder falls in the wide range between that floor and ceiling is not identifiable on-chain. Against the broader EIP-3009 universe (\$772{,}090{,}109.10), x402's facilitator-routed value \$44{,}121{,}383.81 is merely a 5.71\% slice.

\begin{figure}[t]
  \centering
  \begin{minipage}{0.49\columnwidth}\centering
    \begin{tikzpicture}[font=\scriptsize]
    \foreach \v/\yp in {0/0.0, 20/0.657, 40/1.314, 60/1.971}{
      \draw[gray!18] (0,\yp)--(3.0,\yp);\node[font=\tiny, gray!45, anchor=east] at (-0.03,\yp){\v\%};}
    \draw[gray!45] (0,-0.05)--(0,2.36);\draw[gray!45] (3.0,-0.05)--(3.0,2.36);
    \node[anchor=north, font=\tiny] at (0,-0.08){count};\node[anchor=north, font=\tiny] at (3.0,-0.08){value};
    \node[anchor=south, font=\scriptsize] at (1.5,2.4){(a) within x402};
    \draw[c1, very thick] (0,0.697)--(3.0,1.777);\fill[c1](0,0.697)circle(1.5pt);\fill[c1](3.0,1.777)circle(1.5pt);
    \node[c1, anchor=west, font=\tiny] at (0.06,0.697){21.20\%};\node[c1, anchor=east, font=\tiny] at (2.94,1.777){C1~54.08\%};
    \draw[c2, very thick] (0,2.096)--(3.0,0.553);\fill[c2](0,2.096)circle(1.5pt);\fill[c2](3.0,0.553)circle(1.5pt);
    \node[c2, anchor=south west, font=\tiny] at (0.06,2.096){63.78\%};\node[c2, anchor=east, font=\tiny] at (2.94,0.553){C2~16.82\%};
    \draw[c0, very thick] (0,0.494)--(3.0,0.956);\fill[c0](0,0.494)circle(1.5pt);\fill[c0](3.0,0.956)circle(1.5pt);
    \node[c0, anchor=north west, font=\tiny] at (0.06,0.494){15.02\%};\node[c0, anchor=east, font=\tiny] at (2.94,0.956){C3~29.09\%};
  \end{tikzpicture}
  \end{minipage}\hfill
  \begin{minipage}{0.49\columnwidth}\centering
  \begin{tikzpicture}[font=\scriptsize]
    \foreach \v/\yp in {0.1/0.33, 1/0.99, 10/1.65, 100/2.31}{
      \draw[gray!18] (0,\yp)--(3.0,\yp);\node[font=\tiny, gray!45, anchor=east] at (-0.03,\yp){\v\%};}
    \draw[gray!45] (0,0.2)--(0,2.44);\draw[gray!45] (3.0,0.2)--(3.0,2.44);
    \node[anchor=north, font=\tiny] at (0,0.17){count};\node[anchor=north, font=\tiny] at (3.0,0.17){value};
    \node[anchor=south, font=\scriptsize] at (1.5,2.47){(b) EIP-3009 universe};
    \draw[c1, very thick] (0,2.256)--(3.0,1.391);\fill[c1](0,2.256)circle(1.5pt);\fill[c1](3.0,1.391)circle(1.5pt);
    \node[c1, anchor=south west, font=\tiny] at (0.06,2.256){82.82\%};\node[c1, anchor=east, font=\tiny] at (2.94,1.391){C1+C2~4.05\%};
    \draw[c0, very thick] (0,0.074)--(3.0,2.275);\fill[c0](0,0.074)circle(1.5pt);\fill[c0](3.0,2.275)circle(1.5pt);
    \node[c0, anchor=north west, font=\tiny] at (0.06,0.074){0.04\%};\node[c0, anchor=east, font=\tiny] at (2.94,2.275){${>}\$1$k~88.67\%};
  \end{tikzpicture}
  \end{minipage}
 \caption{Count-value structure within x402 by tier (a) and against the EIP-3009 universe (b): share of settlement count (left of each panel) versus share of value (right). }
  \label{fig:concentration}
\end{figure}
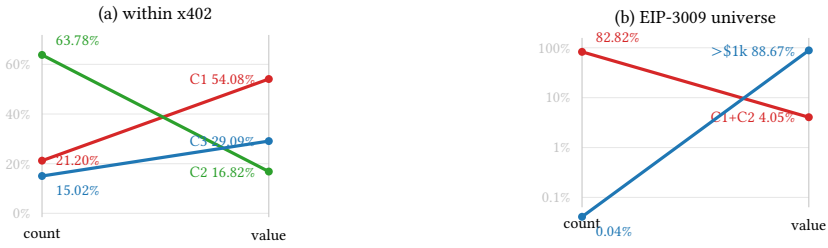

Predominant capital never leaves the operator clusters. Count and value are decoupled; the activity that dominates the count carries little of the value, and vice versa. Specifically, within x402 (Fig.~\ref{fig:concentration}a), the count-dominant internal-settlement tier C2 is 63.78\% of settlements but only 16.82\% of value, while the fictitious self-payment tier C1 is 21.20\% of settlements yet 54.08\% of value. The same decoupling is far starker over the broader EIP-3009 universe (Fig.~\ref{fig:concentration}b): C1+C2 activity is 82.82\% of these settlements but 4.05\% of value, while the 57{,}356 EIP-3009 transfers above \$1{,}000, only 0.04\% of settlements, carry 88.67\% of value (\$684{,}644{,}321.20). 

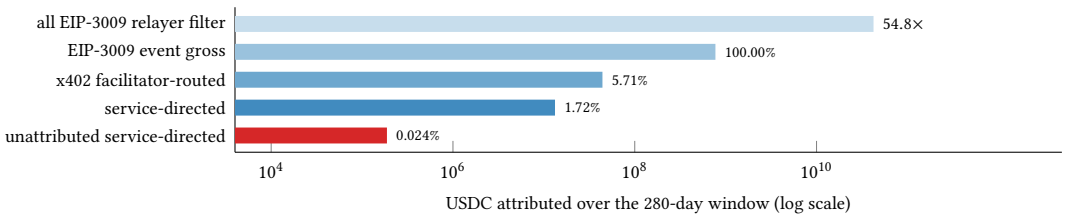
\begin{figure}[t]
  \centering
  \begin{tikzpicture}
  \begin{axis}[
    width=0.9\columnwidth, height=3.5cm, clip=false,
    xmode=log, xmin=4000, xmax=5e12,
    xlabel={USDC attributed over the 280-day window (log scale)},
    xtick={1e4,1e6,1e8,1e10}, xticklabels={$10^4$,$10^6$,$10^8$,$10^{10}$},
    ymin=0.4, ymax=5.6,
    ytick={1,2,3,4,5},
    yticklabels={unattributed service-directed,service-directed,x402 facilitator-routed,{EIP-3009 event gross},{all EIP-3009 relayer filter}},
    axis x line*=bottom, axis y line*=left, ytick style={draw=none},
    tick label style={font=\scriptsize}, label style={font=\scriptsize}]
    \addplot[c0!25, line width=6pt] coordinates {(4000,5) (42348540765,5)};
    \addplot[c0!45, line width=6pt] coordinates {(4000,4) (772090109,4)};
    \addplot[c0!65, line width=6pt] coordinates {(4000,3) (44121384,3)};
    \addplot[c0!85, line width=6pt] coordinates {(4000,2) (13263692,2)};
    \addplot[c1, line width=6pt] coordinates {(4000,1) (187861.35,1)};
    \node[anchor=west, font=\tiny] at (axis cs:42348540765,5) {$54.8\times$};
    \node[anchor=west, font=\tiny] at (axis cs:772090109,4) {100.00\%};
    \node[anchor=west, font=\tiny] at (axis cs:44121384,3) {5.71\%};
    \node[anchor=west, font=\tiny] at (axis cs:13263692,2) {1.72\%};
    \node[anchor=west, font=\tiny] at (axis cs:187861.35,1) {0.024\%};
  \end{axis}
  \end{tikzpicture}
 \caption{Value deflation across identification rules on Base over the same 280-day window. Each rule attributes an orders-of-magnitude-different dollar total. }
  \label{fig:teaser}
\end{figure}

Finally, the x402-related dollar falls off sharply as the counting rule tightens (Fig.~\ref{fig:teaser}). A permissive relayer-EOA filter counts \$42.3B, but it also sweeps in non-x402 treasury and DeFi flow; the raw \texttt{AuthorizationUsed} event gross is \$772{,}090{,}109.10. Restricting to the listed x402 facilitators drops it to \$44{,}121{,}383.81 (5.71\% of the gross), and restricting further to dollars that reach an advertised service drops it to \$13{,}263{,}692 (1.72\%). The unattributed revenue reaching a nameable service is \$187{,}861.35, a demonstrable floor five orders of magnitude below the loosest, though the genuine total could run up to the \$20{,}258{,}746.09 ceiling if the unnameable off-catalog remainder is real. x402 dollar volume is therefore predominantly cluster-internal capital, provably fictitious in the majority, rather than purchasing power for identifiable independent services.

\subsection{The Price of Manufacture: A Subsidized Economy}
\label{sec:manipulability}

Settlement count is an incentivized metric. Public x402 indices such as x402scan rank services by completed-transaction count~\cite{x402scan}, and several ecosystem tokens tie their narrative to settlement throughput~\cite{ping}, so inflating the count carries a standing reward. The fourth link is why the capital scales to this size: each added settlement is almost \textit{free} to the manufacturers. Specifically, a settlement is a gasless signed authorization and the facilitator sponsors the on-chain gas, so the operator's marginal cost of one more settlement is near zero; it fronts only the working capital it cycles, which it then recovers. Reproducing all $S=136{,}708{,}672$ settlements would cost about \$355{,}583 in gas~\footnote{Estimated as the per-settlement \texttt{transferWithAuthorization} gas cost times $S$: from 39 sampled relayer receipts the mean is 86{,}288 gas units at 0.0098 gwei, i.e.\ $8.46\times10^{-7}$ ETH per settlement. Over $S=136{,}708{,}672$ settlements and Base ETH prices across the window (\$1{,}736--\$4{,}349) this gives a central \$355{,}583, with a \$96k--\$554k band spanning the cheap-batch and busy-day gas rates. A facilitator-by-month stratified re-estimate, sampling gas receipts within each of 206 (facilitator, month) strata and volume-weighting the per-settlement gas by each stratum's settlement share (612 receipts), gives \$142{,}996, within this band; batching is negligible (1.015 settlements per transaction), so the difference is that Base gas ran cheaper in the high-volume 2025-11 and 2025-12 campaign months than on the single June sampling day. }, while paid by the relayer, i.e. the facilitator. The clearest case is the self-payment fleets: the 2025-11 fleet settled 27{,}089{,}029 times from itself to itself, so no capital ever left and its net outlay was nil by construction. Moreover, fan-in-with-sweep recovers the principal while spending a tiny part of it: lnpay's hub absorbed 31{,}551{,}433 settlements (23.08\% of $S$), seeded with \$415{,}765 of which \$408{,}848 (98.34\%) was swept back to the funder, a true net cost of \$6{,}917 for tens of millions of settlements.
Because the operators account for 67.76\% of settlements, they absorb \$240{,}943 of the \$355{,}583 in gas, so the public-goods subsidy meant to bootstrap real usage funds mostly cluster-internal traffic. The subsidy also supplies its own test: when the dominant facilitator's per-settlement fee took effect on 2026-01-01~\cite{cdpfee}, ending the free margin, the series registered its largest sustained fall (\S\ref{sec:dynamics}), volume behaving as if priced at the subsidy.

\subsection{Dynamics: Operator-Driven}
\label{sec:dynamics}
The last link is the arc the economy traces over the window, how it forms, expands, and subsides. Demand-driven growth would track exogenous events and widen the user base. In this series, every large movement is dated to an operator's own action, the growth is intensive (more settlements per existing payer) rather than extensive (more distinct payers), and the same structure recurs whenever an operator switches a campaign on.

\subsubsection{Inflection Points Align with Operator Actions}
\begin{figure}[t]
  \centering
  \begin{tikzpicture}
  \begin{axis}[width=\columnwidth, height=4cm, ymode=log, ylabel={settlements / day},
    xmin=0, xmax=279, ymin=999, ymax=10000001, xtick={14,75,137,196,257},
    xticklabels={2025-10,2025-12,2026-02,2026-04,2026-06}, ymajorgrids]
    \addplot[blue!65!black, line width=0.5pt] table[x=x, y=settlements]{fig/adoption.dat};
    \draw[black!55, dotted, line width=0.7pt] (axis cs:38,1000) -- (axis cs:38,5000000);
    \draw[red!70!black, dashed, line width=0.5pt] (axis cs:57,1000) -- (axis cs:57,5000000);
    \draw[red!70!black, dashed, line width=0.5pt] (axis cs:72,1000) -- (axis cs:72,5000000);
    \draw[red!70!black, dashed, line width=0.5pt] (axis cs:82,1000) -- (axis cs:82,5000000);
    \draw[black!55, dotted, line width=0.7pt] (axis cs:106,1000) -- (axis cs:106,5000000);
    \draw[red!70!black, dashed, line width=0.5pt] (axis cs:265,1000) -- (axis cs:265,5000000);
    \node[font=\tiny, fill=white, inner sep=0.4pt] at (axis cs:38,7000000) {1};
    \node[font=\tiny, fill=white, inner sep=0.4pt] at (axis cs:57,7000000) {2};
    \node[font=\tiny, fill=white, inner sep=0.4pt] at (axis cs:72,7000000) {3};
    \node[font=\tiny, fill=white, inner sep=0.4pt] at (axis cs:82,7000000) {4};
    \node[font=\tiny, fill=white, inner sep=0.4pt] at (axis cs:106,7000000) {5};
    \node[font=\tiny, fill=white, inner sep=0.4pt] at (axis cs:265,7000000) {6};
  \end{axis}
  \end{tikzpicture}
 \caption{Daily settlements (log scale). The six numbered movements are the largest in the series, dashed for operator actions and dotted for external events: (1) 2025-10-25, $+1{,}460{,}132$, the October take-off at the PING pay-to-mint launch (external); (2) 2025-11-13, $+1{,}958{,}337$, the 2025-11 self-payment fleet switching on, the sharpest jump in the window; (3) 2025-11-28, $-1{,}791{,}905$, the same fleet switching off as its three wallets cease together; (4) 2025-12-08, $+646{,}615$, the December fan-in carried by lnpay and t54; (5) 2026-01-01, $-683{,}998$, the collapse as the facilitator's per-settlement fee takes effect (external); (6) 2026-06-09, $+395{,}309$, a single verified payer-payee pair switching on. }
  \label{fig:adoption}
\end{figure}
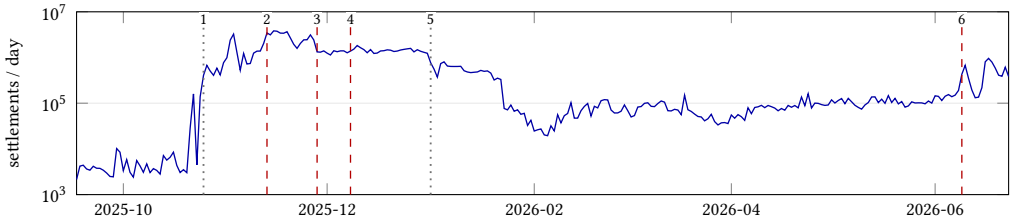

We locate where the daily settlement series changes and ask, for each change, whether it is a sharp switch-on or a gradual regime shift, using two complementary estimators. The first is a segmented regression that lets both the level and the slope of the trend change at candidate break dates, with Newey-West HAC errors~\cite{neweywest} and Holm~\cite{holm} and Benjamini-Hochberg~\cite{bh} correction over the battery of principal tests. It is a global fit, sensitive to sustained changes in level or slope, but it reads each date net of the whole window's trend. Its estimate at one date can therefore be pulled by breaks elsewhere, and a sharp jump that later reverts need not register. The second is a local regression-discontinuity in time~\cite{rdit}: inside a symmetric $\pm21$-day window at each dated cutoff $c$ we fit $y_t = \alpha + \beta\,(t-c) + \tau\,\mathbf{1}[t\ge c] + \gamma\,(t-c)\mathbf{1}[t\ge c]$ with HAC errors, so $\tau$ is the level jump at the boundary. We then re-run the identical estimator at every interior day at least 14 days from any dated event (159 placebo cutoffs) to build a null distribution against which the largest jumps stand out as outliers (Fig.~\ref{fig:adoption}). The local test isolates the sharp single-day discontinuities the global fit blurs, while the slow slope changes a short window cannot resolve remain the segmented model's job; together they separate a switch-on from a gradual composition change. Each locates where the series shifts, and we read the dates against the operators of \S\ref{sec:rq2} and the ecosystem timeline.

Ranked by magnitude, the largest movements in the series are overwhelmingly the operators' own campaigns turning on and off (Fig.~\ref{fig:adoption}). The two biggest are a single campaign: the 2025-11 self-payment fleet switches on, the sharpest local discontinuity in the window; and switches off two weeks later, when its three wallets cease within the same two-second window (\S\ref{sec:rq2}). The largest sustained fall, the January collapse, coincides to the day with the facilitator's per-settlement fee taking effect (effective 2026-01-01~\cite{cdpfee}), consistent with the subsidized economy of \S\ref{sec:manipulability}. The fall continues into late January as the fleets wind down, though the fee and the shutdown fall within days and we do not separate them. The last movement, the June rebound, is a single verified payer-payee pair switching on. Only one large movement has an external catalyst: the October take-off, which tracks the PING pay-to-mint launch~\cite{ping}, incentive-driven token-minting rather than service demand. Against all of this, the x402 v2 release (announced 2025-12-11~\cite{x402v2}), lands on the December plateau and moves the series not at all. The turning points are authored by operators and one facilitator decision; the ecosystem's headline release leaves no mark.

\subsubsection{Campaigns Rotate over a Narrow Base}
Operators rotate campaigns, and the concentration of settlement activity across payers stays extreme in every quarter. The Gini of the per-payer settlement-count distribution is 0.99 in the October--December 2025 campaign quarter, eases only to 0.93 in the January--March 2026 post-collapse quarter, and returns to 0.97 in April--June 2026. A coefficient this high means a small set of payers accounts for almost all settlements. It never falls below 0.93, not even in the quiet quarter after the largest fleets have stopped, which is the signature of a rotating set of narrow campaigns rather than a broadening base of independent payers. A widening user population would drag the coefficient toward the moderate values a diverse base produces, well below the 0.9 range, and no quarter moves in that direction. The campaigns are also transient. Once they subside, the January--June 2026 tail runs at about 145{,}500 settlements per day (25{,}319{,}239 in total, 18.52\% of $S$), an order of magnitude below the November--December peaks.

\subsubsection{Growth Is Intensive, Not Extensive}
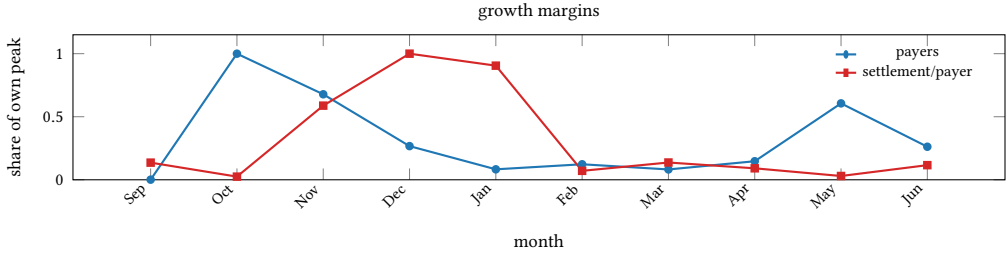
\begin{figure}[t]
  \centering
  \begin{tikzpicture}
  \begin{axis}[width=\linewidth, height=3.5cm, ymin=0, ymax=1.15,
    symbolic x coords={Sep,Oct,Nov,Dec,Jan,Feb,Mar,Apr,May,Jun}, xtick=data,
    xlabel={month}, ylabel={share of own peak}, xlabel style={font=\scriptsize}, ylabel style={font=\scriptsize}, tick label style={font=\tiny},
    x tick label style={font=\tiny, rotate=45, anchor=east}, ytick={0,0.5,1.0},
    title={growth margins}, title style={font=\scriptsize, yshift=-1ex},
    legend style={font=\tiny, at={(0.98,0.98)}, anchor=north east, legend columns=1, row sep=-2pt, draw=none, fill=none}, legend image post style={xscale=0.45}]
    \addplot[c0, thick, mark=*, mark size=1.2pt] coordinates {(Sep,0.001)(Oct,1.0)(Nov,0.678)(Dec,0.267)(Jan,0.083)(Feb,0.123)(Mar,0.082)(Apr,0.147)(May,0.606)(Jun,0.262)}; \addlegendentry{payers}
    \addplot[c1, thick, mark=square*, mark size=1.2pt] coordinates {(Sep,0.136)(Oct,0.025)(Nov,0.588)(Dec,1.0)(Jan,0.905)(Feb,0.071)(Mar,0.137)(Apr,0.091)(May,0.030)(Jun,0.116)}; \addlegendentry{settlement/payer}
  \end{axis}
  \end{tikzpicture}
  \caption{Growth is on the intensive margin, not the extensive one (each monthly series normalized to its own peak): the settlement explosion (intensive, settlements per payer, peaking in December) coincides with a \emph{declining} distinct-payer base (extensive, peaking in October), i.e.\ fewer wallets each settling more.}
  \label{fig:growthmargins}
\end{figure}

The volume curve grew on the wrong margin for demand-driven adoption. We decompose monthly settlements into the extensive margin, the count of distinct active payers; and the intensive margin, settlements per active payer. The November--December explosion to 64{,}264{,}274 and 43{,}037{,}136 settlements coincided with a \emph{shrinking} payer base: from 175{,}234 distinct payers in October down to 118{,}764 in November, 46{,}791 in December, and a trough of 14{,}581 in January. Settlements per payer meanwhile rose from 23 in October to 541 in November, 920 in December, and 832 in January (Fig.~\ref{fig:growthmargins}). Demand-driven adoption grows the extensive margin (more distinct users); x402's growth is intensive, and the distinct-payer count is never seen growing monotonically across the window.

\subsubsection{Advertised Supply and Settled Value Are Decoupled}
\begin{figure}[t]
  \centering
  \begin{tikzpicture}
  \begin{loglogaxis}[width=\linewidth, height=4cm,
    xlabel={advertised Base resources}, ylabel={USD earned},
    xlabel style={font=\scriptsize}, ylabel style={font=\scriptsize},
    xmin=0.8, xmax=16000, ymin=0.1, ymax=1e8, tick label style={font=\tiny},
    title={supply vs demand}, title style={font=\scriptsize, yshift=-1ex},
    legend style={font=\tiny, at={(0.97,0.97)}, anchor=north east, legend columns=1, row sep=-1.5pt, draw=none, fill=none}]
    \addplot[only marks, mark=*, black!35, mark size=0.5pt] table {fig/sd_advertised.dat}; \addlegendentry{advertised}
    \addplot[only marks, mark=square*, c1, mark size=1.1pt] table {fig/sd_factory.dat}; \addlegendentry{template}
    \addplot[only marks, mark=triangle*, c0, mark size=1.6pt] table {fig/sd_operator.dat}; \addlegendentry{operator}
  \end{loglogaxis}
  \end{tikzpicture}
  \caption{Supply and demand are inflated by different actors (log-log; each Base catalog payee that ever settled is a point): templates flood the catalog yet take in almost nothing, the operator fleets take in the dollars yet advertise next to nothing, and the high-supply, high-demand corner is empty.}
  \label{fig:supplydemand}
\end{figure}
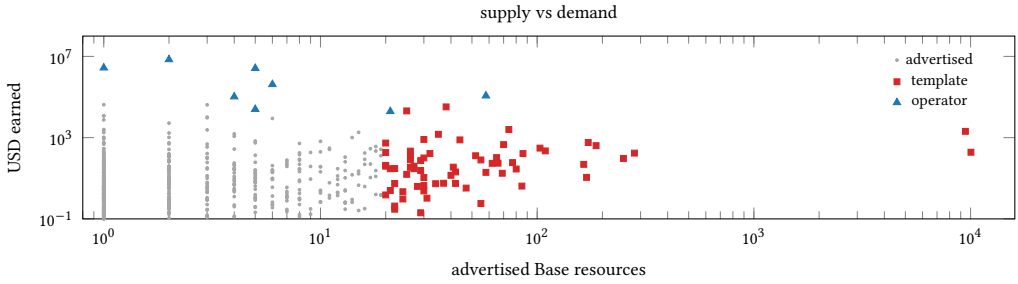

The two quantities, that how many services are advertised and how many transactions settle, come apart: a large advertised-resource count and a large settled value are held by different payees and seldom by the same one. On the supply side, two payTo addresses account for 77.48\% of all advertised Base resources (one with 10{,}035 listings, another with 9{,}461), yet these two settle only about \$2{,}234 between them, so a high listing count need not correspond to appreciable settled value. On the demand side, the payees that take in the dollars are a small set of hubs (lnpay, t54, and the 2025-11 self-pay run) that advertise few or zero catalog resources. Plotting every Base catalog payee that ever settled on the two axes shows the two seldom coincide (Fig.~\ref{fig:supplydemand}): of the ten payees advertising at least 100 resources none earns even \$100{,}000, while the six payees that earn more than \$100{,}000 each advertise at most 58 resources. The high-supply, high-demand corner is therefore empty. 

\takeaway{RQ3}{The five properties are one predictable equilibrium: settlement count is both \emph{rewarded} (leaderboards, volume-linked tokens) and, with gas sponsored, nearly \emph{free}, so a cost-minimizing operator manufactures exactly what we observe. Yet the on-chain marker cannot itself separate this from genuine demand; only orthogonal axes, e.g., timing, funding, survival, do, and the economy they reveal is small. Settlement count is a Goodhart metric: subsidized and rewarded, it measures the incentive to transact, not the adoption it is taken to prove.}

\section{Discussion}
\label{sec:limitations}

\subsection{Validity and Limitations}

\paragraph{What can and cannot be observed onchain.} The chain records the settlements; whether an off-chain economic event, an HTTP request served or a resource delivered, stands behind a given settlement is not observable from on-chain transfers alone. The verdicts above therefore assert only what settlement traces can prove: that the flagged settlements move value among wallets in one funding-linked cluster, funded on-chain by a recirculating pool, and so do not constitute payments between independent parties and cannot be read as that many independent economic interactions. This holds under every reading of the off-chain world. Even if such a cluster internally meters a real off-chain workload, its settlements remain internal settlement throughput, just as a payment processor's (e.g., Stripe's) internal ledger entries are not counted as merchant adoption. We claim neither that a flagged operator has no off-chain business, nor anything about intent.

\paragraph{Robustness of the operator-internal classification.} Only the fictitious tier (C1, 21.20\% of $S$) is proven manufactured, and the internal-settlement verdict rests on neither the lone shared-funder signal nor the named operator anchors. The largest internal-settlement clusters each carry two or more independent signals, and granting every singly-corroborated cluster the benefit of the doubt still bounds the operator-internal share within $[72.38\%, 84.98\%]$; striking the single largest cluster outright leaves it at 61.90\%. The details are in Appendix~\ref{app:sybilbaseline}.

\paragraph{Coverage and scope.} On Base the EIP-3009 authorized transfer is the sole in-spec settlement path, and our rule intersects that event with a two-source facilitator allowlist (30 facilitators, 129 relayer EOAs; Appendix~\ref{app:allowlist}), a deliberate bound that admits neither the facilitators' non-x402 flow nor the full EIP-3009 event universe; Solana, lacking the event, is only a coarser confirmed-facilitator upper bound and is never merged with Base. The advertised supply and the \$187{,}861.35 that reaches a nameable service are both defined against one discovery catalog, the CDP Bazaar, so services discoverable only off-catalog are invisible to the supply funnel and could raise that floor, which is why we report the genuine economy as a bound rather than a point estimate. The HTTP liveness probe and the catalog are point-in-time snapshots, and the fixed 280-day window is right-censored at its boundary, so late-arriving activity is not tracked.

\subsection{Implications}
The structure that makes settlement count manufacturable, an emit-cheap action tied to a standing reward, is not specific to x402: any on-chain adoption metric that is subsidized to produce and linked to leaderboards or volume-indexed tokens invites the same inflation. Reading such a count as adoption requires evidence orthogonal to the emitting act, whether value reaches a party outside the transactor's control, or an off-chain attestation that a resource was delivered; absent that, the count certifies only that emitting was cheap and rewarded. The subsidy itself illustrates the mismatch here, since the facilitator gas meant to bootstrap real usage funds mostly cluster-internal traffic, the operators absorbing \$240{,}943 of the \$355{,}583 total.

\subsection{Reproducibility and Ethics}

\paragraph{Reproducibility and artifact.} The identification, role resolution, clustering, and classification run over the fixed 280-day window from on-chain data and a point-in-time catalog snapshot. Upon publication we will release the full pipeline and dataset, so that new windows, facilitators, assets, and chains can be re-measured. 

\paragraph{Ethics.} The study uses only public on-chain data and a public discovery catalog and makes no claim about any party's intent or off-chain business. The HTTP liveness probe (\S\ref{sec:supply}) was a single GET per host to an already-advertised endpoint, performed no payment, and was rate-limited, so it neither transacts nor stresses a service. Operator names are read from the operators' own public self-advertisement in the catalog, a hub listed under its own domain, not from any de-anonymization of individuals. We name only such self-advertised operator handles, and do not re-identify the controllers behind pseudonymous addresses that are already public.

\section{Related Work}
\label{sec:related}

\paragraph{Blockchain measurement, fund flows, and clustering.} Population-scale blockchain measurement has an established lineage, from Bitcoin transaction-graph analysis that recovered entity structure from raw transfers~\cite{fistful} and ERC20 token-network studies~\cite{victorluders} to reproducible analytical platforms~\cite{blocksci}. Recent work censuses emerging on-chain phenomena: illicit money-flow tracing~\cite{mftracer}, cross-chain MEV~\cite{crosschainarb}, transaction matching across heterogeneous chains~\cite{jigsaw}, phishing contracts~\cite{phishingcontracts}, amplification attacks~\cite{blockchainamp}, misbehavior on a major chain~\cite{eosio}, and counterfeit tokens~\cite{counterfeit}. A security-venue line bears directly on our attribution and validation: address-cluster validation and expansion~\cite{kappos}, methodology-sensitive revenue estimation~\cite{gomez}, token-scam censuses~\cite{tokenspammers}, scam-token detection~\cite{tokenscout}, and attribution audits~\cite{ghostclusters}.

\paragraph{Detecting manufactured activity and measuring adoption.} Telling genuine activity from manufactured draws on wash-trading detection on centralized and decentralized markets~\cite{washtrade,dexwash,nftwash}, behavioral bot detection on Ethereum~\cite{financialbots}, Sybil and airdrop-farming detection~\cite{trustasybil}, weak supervision for scarce labels~\cite{snorkel}, and ground-truth-by-construction in systems measurement~\cite{spamalytics,canarytrap}; \S\ref{sec:genuine} dissects the criteria these lines established and shows which of them survive an agentic-payment setting. For adoption itself, Internet measurement has long separated advertised from effective deployment when tracking protocol rollout, charting HTTPS and certificate uptake~\cite{httpsadoption,letsencrypt}, new transport and TLS versions~\cite{quic,tls13}, and the deployment of LLM services in the wild~\cite{llmwild}. A causal-inference line attributes on-chain change to dated events with quasi-experimental designs around upgrades and incentive launches~\cite{l2adoption,decentralizationquasi} and before-after comparisons across a consensus change~\cite{ethpos}.

\paragraph{Agentic-payment security and identity.} A parallel line studies the security of x402-style payment systems, enumerating vulnerability classes and attack feasibility, and the registration of agent identity~\cite{freeride,erc8004}. It is orthogonal to the axis measured here. Identity registration counts registrations, and the security work establishes that manipulation is feasible; neither measures settlement-level adoption nor how much observed activity is genuine. To our knowledge, this is the first population-scale measurement of agentic-payment adoption and of how much of it reflects genuine demand. Prior quantifications of x402 come from vendor dashboards and chain-analytics reports, none of them peer-reviewed; industry analyses have observed self-dealing and reconciled divergent cross-source volume figures~\cite{a16zaudit}, but qualitatively and without a reproducible identification rule or a validated, population-scale authenticity decomposition. Those trackers moreover disagree with one another by an order of magnitude on 30-day x402 volume over identical periods~\cite{a16zaudit}, which is itself evidence that adoption claims need an explicit identification rule and counting unit.

\section{Conclusion}
\label{sec:conclusion}

Measured at population scale, agentic commerce on x402 is far smaller and more concentrated than its settlement count implies. Over a fixed 280-day window we census 136{,}708{,}672 Base settlements worth \$44{,}121{,}383.81; sorting each by what its on-chain trace proves, 84.98\% of the count is operator-internal, and what could be genuine is only bounded, from the \$187{,}861.35 that demonstrably reaches a nameable service up to the \$20{,}258{,}746.09 (45.92\% of value) not provably manufactured. What fills the count is a coherent operator-driven economy, star-shaped, machine-timed, and so gas-subsidized that reproducing it costs about \$355{,}583 while paid by facilitators. Settlement count is thus a Goodhart metric, measuring the incentive to transact rather than the adoption it is read as proving. We will release the pipeline and dataset upon publication. Our findings do not dispute that x402 may support real activity; they show its counts cannot be read as independent adoption.

\bibliographystyle{ACM-Reference-Format}
\bibliography{refs}

\appendix

\section{Facilitator Allowlist}
\label{app:allowlist}
The x402 ecosystem is permissionless and has no authoritative registry, so we anchor the allowlist on a commercial on-chain index that labels facilitators on its own maintained address set (Allium's \texttt{base.agents.x402\_transfers}) and cross-validate it against an independently maintained open-source community curation (the x402scan \texttt{facilitators/lists/all.ts}). Over the window the index labels 30 facilitators (Table~\ref{tab:allowlist}), which resolve to 129 relayer EOAs (a single facilitator can operate several; the per-facilitator EOA counts sum to the 129 distinct addresses with no sharing) and reproduce the principal census exactly: 136{,}708{,}672 settlements and \$44{,}121{,}383.81. The community curation lists 29 facilitators, every one of which the index also labels, so the two sources agree on 29 of 30 entries and their union is 30. The single index-only facilitator, Nuwa Protocol, carries 13{,}236 settlements, 0.0097\% of the census, so the cross-source disagreement is immaterial to every reported quantity. The 129 relayer EOAs are a strict subset of the 5{,}587 relayers that emit an \texttt{AuthorizationUsed} event in the window.
  
\begin{table}[t]
  \caption{The 30 facilitators of the x402 allowlist from Allium, as of June 2026.
  Comparison with x402scan shows that 29 of the
  30 facilitators are shared; only Nuwa Protocol is absent from
  the x402scan allowlist. Per-facilitator USD entries are independently rounded to the cent and may sum a cent or two off the exact total.}
  \label{tab:allowlist}
  \footnotesize
  \setlength{\tabcolsep}{4pt}

  \begin{tabular}{lrrr@{\hspace{3em}}lrrr}
  \toprule
  Facilitator & Settlements & EOAs & Gross USD &
  Facilitator & Settlements & EOAs & Gross USD \\
  \midrule

  Coinbase          & 86,859,685 & 40 & 27,892,703.88 &
  Dexter            &    40,248 & 2 &     2,471.27 \\

  PayAI             & 23,521,216 & 15 &  4,372,475.07 &
  Polymer           &    38,273 & 1 &     5,148.82 \\

  Daydreams         & 11,810,802 &  2 &  2,757,672.98 &
  Cascade           &    21,562 & 1 &        81.85 \\

  Heurist           &  7,964,123 &  9 &     30,075.60 &
  Corbits           &    19,777 & 1 &       304.57 \\

  Virtuals Protocol &  1,989,097 &  1 &  4,339,477.70 &
  Primer            &    19,342 & 1 &     3,862.57 \\

  FluxA             &    962,466 &  6 &    103,861.02 &
  Mogami            &    18,636 & 1 &   310,273.55 \\

  X402rs            &    697,618 &  6 &    468,665.61 &
  OpenFacilitator   &    14,509 & 1 &    30,909.97 \\

  OpenX402          &    560,994 &  2 &    174,171.66 &
  xEcho             &    14,074 & 1 &     2,847.30 \\

  AnySpend          &    514,164 &  1 &    129,109.25 &
  Nuwa Protocol     &    13,236 & 3 &        94.24 \\

  CodeNut           &    477,859 &  4 &    110,005.03 &
  x402 Jobs         &     7,599 & 1 &       366.25 \\

  Questflow         &    473,868 & 10 &     10,017.70 &
  Ultravioleta DAO  &     7,448 & 1 &       490.86 \\

  AurraCloud        &    235,444 &  3 &     53,540.84 &
  Treasure          &     2,344 & 1 &       398.22 \\

  Thirdweb          &    220,302 & 10 &    118,721.80 &
  Openmid           &       234 & 1 &        34.68 \\

  Meridian          &    114,180 &  1 &  2,988,335.45 &
  Bitrefill         &        95 & 1 &       302.69 \\

  RelAI             &     89,447 &  1 &    214,951.64 &
  402104            &        30 & 1 &        11.75 \\

  \midrule
  \multicolumn{8}{c}{\textbf{Total (30 facilitators):}
  136,708,672 settlements, 129 EOAs, USD 44,121,383.81} \\
  \bottomrule
  \end{tabular}
\end{table}

On Solana the same two curations agree exactly. The commercial index labels 15 facilitators in its Solana x402 table (Allium's \texttt{solana.agents.x402\_transfers}, Table~\ref{tab:allowlist-solana}), and the open-source community curation (x402scan) independently carries a Solana fee-payer address for the same 15 facilitators and for no others. This 15-of-15 agreement is tighter than the 29-of-30 on Base. The crucial difference from Base is that Solana exposes no clean event marker, so this fee-payer-signer set constitutes the entire identification. The resulting 49{,}477{,}928-settlement count is therefore a confirmed-facilitator upper bound, dominated by a single facilitator (Dexter, 67.72\% of the count). A per-transaction \texttt{TransferChecked}-shape validity check bounds the contamination of that coarser unit (\S\ref{sec:idvalidity}): only 0.307\% of settlements carry the facilitator address as payer or recipient and 0.0056\% are self-transfers, so it is tight rather than loose.

\begin{table}[t]
    \caption{The 15 facilitators of the Solana x402 set from Allium, as of June 2026. Comparison with x402scan shows that all 15 facilitators are shared. Unlike Base, Solana exposes no clean event marker, so this set is the whole identification and the count is a confirmed-facilitator upper bound. Per-facilitator USD entries are independently rounded to the cent and may sum a cent or two off the exact total.}
    \label{tab:allowlist-solana}
    \footnotesize
    \setlength{\tabcolsep}{4pt}

    \begin{tabular}{lrr@{\hspace{3em}}lrr}
    \toprule
    Facilitator & Settlements & Gross USD &
    Facilitator & Settlements & Gross USD \\
    \midrule

    Dexter           & 33,504,613 & 5,842,237.77 &
    x402 Jobs        &    15,607 &     2,178.33 \\

    PayAI            & 14,572,595 & 2,488,399.11 &
    Cascade          &     8,556 &        90.45 \\

    Coinbase         &    830,290 &   287,414.85 &
    Ultravioleta DAO &     1,854 &        21.28 \\

    Corbits          &    212,527 &     1,329.53 &
    AurraCloud       &       543 &         0.06 \\

    RelAI            &    191,376 &   283,335.40 &
    AnySpend         &        68 &         7.83 \\

    OpenX402         &    103,560 &    85,599.15 &
    CodeNut          &        63 &         7.76 \\

    OpenFacilitator  &     18,691 &    42,076.38 &
    Bitrefill        &         3 &         6.00 \\

    Daydreams        &     17,582 &         2.20 &
                     &           &              \\

    \midrule
    \multicolumn{6}{c}{\textbf{Total (15 facilitators):}
    49,477,928 settlements, USD 9,032,706.12} \\
    \bottomrule
    \end{tabular}
\end{table}

\section{Grading the Internal-Settlement Evidence}
\label{app:sybilbaseline}

The graph assigns 84.98\% of $S$ to the fictitious and internal-settlement tiers (C1 and C2, \S\ref{sec:method-graph}), resting in part on named recipient hubs (lnpay, t54) and their seed funders. 
A shared funder does not by itself show that a cluster's settlements are internal rather than between independent parties: a custodial agent-wallet platform, an exchange, or a bridge also seeds many wallets from one address (\S\ref{sec:genuine}). Two design choices and a grading step keep C2 from resting on that lone signal.

First, the merge rule requires a \emph{bespoke} funder, not merely a shared one. A funder qualifies only when its out-neighborhood is near-confined to a single operator's fleet and hub. Exchanges, bridges, routers, relayers, and smart-wallet factories, the high-degree shared addresses whose recipients span the whole chain, are held out as opaque boundaries and never merge the wallets they touch (\S\ref{sec:method-graph}). This is a structural negative control: because a custodian's fundees trace to no single hub, the rule cannot fold an exchange's customers into one operator. The 2025-11 self-payment wallets show it on real data, first-funded from two exchange wallets (Binance and Gate) yet not merged with the rest of those exchanges' fundees, and classified C1 only by their self-payment endpoint.

Second, we grade C2 by how many \emph{independent} signals corroborate that each cluster's settlements are internal, requiring more than shared funding alone. The C2 mass is concentrated in a few clusters that each carry two or more mutually independent signals: the three largest, lnpay, t54/lucyos, and ainalyst, together carry 69{,}974{,}099 settlements, 51.18\% of $S$ and 80.2\% of the C2 mass, and none rests on shared funding alone. The lnpay cluster adds sweep-closure (98.34\% of hub inflow returned to its funder) and a fleet whose capital is 99.98\% internal; t54/lucyos adds synchronized activation (70.33\% of payers first settling on one day) and a hub self-advertised under its own domain; ainalyst adds a 16-second fund-and-repay sweep. Sweep-closure, synchronized activation, and a self-advertised hub are each a distinct on-chain fact a shared funder does not imply.

Third, we carry the residual uncertainty into a bound, not a point claim. Consider every C2 cluster that rests on shared funding \emph{without} a second corroborating signal, at most the 12.60\% of $S$ outside the multiply-corroborated clusters. If all of it is granted the benefit of the doubt and reclassified as possibly independent, the operator-internal share falls from 84.98\% to 72.38\% (C1's 21.20\% plus the multiply-corroborated C2's 51.18\%). The operator-internal share is therefore bounded in $[72.38\%, 84.98\%]$, and even its floor leaves the possibly independent remainder a minority of the count. The bound is also insensitive to any single attribution: striking the largest cluster (lnpay, 23.08\% of $S$) from the operator-internal mass outright still leaves it at 61.90\%, and striking t54/lucyos (its cluster's 21.61\%) instead leaves 63.37\%.

\section{Silver-Set Annotation Protocol and the C3 Residual Model}
\label{app:goldset}
This appendix expands the C3 residual model, whose reported share we treat as an estimate of the otherwise-unresolved tail rather than a validation of genuineness. Its scope is exactly that tail: none of the figures below enter the headline C1 and C2 shares or the genuine-economy bound of \S\ref{sec:residual}, which rest on the structural verdicts alone.

\paragraph{Sampling and annotation.} The silver set is a stratified random sample of 500 payers drawn to span the full settlement-volume range, so that low- and high-activity payers are both represented. Each payer was independently adjudicated by 7 annotators under a three-label codebook, \emph{organic} (independent demand), \emph{non-organic} (operator-generated or otherwise manufactured), or \emph{ambiguous}, with the payer's on-chain behavioral summary as the evidence shown. We take the per-payer label by majority and keep the 438 payers with a decisive (non-ambiguous) consensus as the labeled set; inter-annotator agreement is Fleiss $\kappa = 0.651$ over the three classes (consensus 280 non-organic, 158 organic, 62 ambiguous).

\paragraph{Features and models.} Each payer is summarized by the 9 behavioral features, computed from its settlements alone: settlement count and value, average ticket, distinct payees, active days, calendar span, daily-count variability, peak single-day count, and signer-wallet nonce. The features deliberately exclude the funding- and sweep-graph signals that define C1 and C2, so that agreement between the behavioral model and the structural tiers is earned rather than built in. We fit three model families, Logistic Regression, Random Forest, and Gradient Boosting, and report their cross-validated non-organic $F_1$, balanced accuracy, and per-class precision and recall in Table~\ref{tab:classifier}; the three agree, so the result does not hinge on a single model choice; we avoid deep learning, which would overfit a labeled set this size and, on tabular data of this shape, does not outperform tree ensembles~\cite{grinsztajn2022,shwartzziv2022}.
\begin{table}[t]
    \caption{The payer-behavioral model against the 7-annotator human silver set. Three model families are reported and agree, so the result is model-agnostic.}
    \label{tab:classifier}
    \centering
    \small
    \begin{tabular}{lrrr}
    \toprule
    & Logistic Regression & Random Forest & Gradient Boosting \\
    \midrule
    $F_1$ (non-organic)     & 0.864 & 0.849 & 0.848 \\
    balanced accuracy       & 0.826 & 0.801 & 0.789 \\
    precision (non-organic) & 0.887 & 0.863 & 0.847 \\
    recall (non-organic)    & 0.843 & 0.836 & 0.850 \\
    \bottomrule
  \end{tabular}
\end{table}

\paragraph{Application and its limits.} Applied to the 457{,}400-payer C3 tail, the three families flag 87.82\% of unclustered payers as manufactured-behaving (95\% CI [87.73\%, 87.92\%]; 84.91\% flagged by all three families, 9.67\% held organic by all three). We read this as a soft estimate for three reasons a stricter external validation would have to remove. First, the annotators judged from behavioral summaries that overlap the model's features, so a stereotype that reads machine-like but legitimate agent traffic as manufactured would be inherited by any model trained on the labels. Second, the three families share both the labels and correlated features, so their agreement is only a consistency check. Third, genuine agent demand and manufactured traffic share automation, periodicity, and dust-sized tickets (\S\ref{sec:genuine}), so behavior alone cannot certify genuineness. For these reasons the estimate is confined to bounding the C3 tail and is kept out of every headline quantity. The codebook, the annotation interface, and the per-fold evaluation are part of the released artifact.

\end{document}